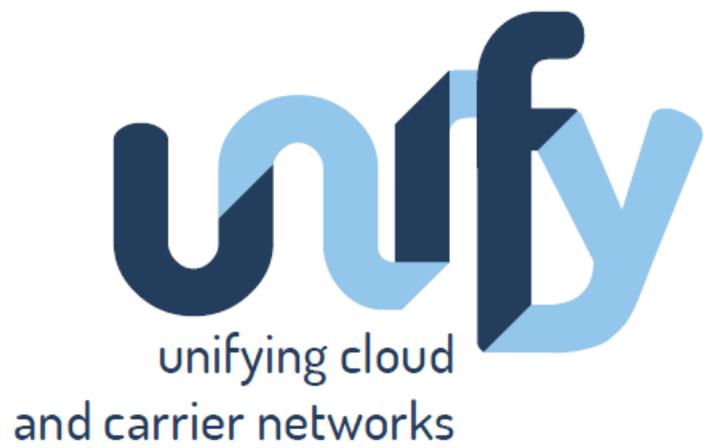

# Deliverable D4.1

Initial requirements for the SP-DevOps concept, Universal Node capabilities and proposed tools

| | |
|---|---|
| Dissemination level | PU |
| Version | 1.1 |
| Due date | 30.06.2014 |
| Version date | 10.02.2015 |

This project is co-funded by the European Union

## Document information

### Editors and Authors:

Editors: Wolfgang John and Catalin Meirosu (EAB)

Contributing Partners and Authors:

| ACREO | Pontus Sköldström |
|---|---|
| BME | Felician Nemeth, Andras Gulyas |
| DTAG | Mario Kind |
| EAB | Wolfgang John, Catalin Meirosu |
| iMinds | Sachin Sharma |
| OTE | Ioanna Papafili, George Agapiou |
| POLITO | Guido Marchetto, Riccardo Sisto |
| SICS | Rebecca Steinert, Per Kreuger, Henrik Abrahamsson |
| TI | Antonio Manzalini |
| TUB | Nadi Sarrar |

### Project Coordinator

Dr. András Császár

Ericsson Magyarország Kommunikációs Rendszerek Kft. (ETH) AB

KONYVES KALMAN KORUT 11 B EP

1097 BUDAPEST, HUNGARY

Fax: +36 (1) 437-7467

Email: andras.csaszar@ericsson.com

### Project funding

7th Framework Programme

FP7-ICT-2013-11

Collaborative project

Grant Agreement No. 619609

### Legal Disclaimer

# Table of contents

# Summary

Management of telecommunication networks is complex and costly. In a recent survey [1], 80% of the participating telecom operators indicated that operational excellence is the most important area for improvements. This need for operational excellence together with the upcoming demand for high velocity in service deployment calls for novel concepts, which evolve beyond existing workflows and processes as defined in current frameworks based on historical best-practices (e.g. eTOM and ITIL). We present two examples of current operational practices that reflect the overhead incurred by operators when introducing a new MPLS VPN service or when operating network middle boxes of types which are expected to be replaced with virtual functions running on the UNIFY production environment.

As part of the UNIFY vision, networking and service functions will be virtualized on commodity hardware, and thus treated as software applications. As outlined in D2.1, services described as service graphs (an association of Virtual Network Functions – VNFs - and their interconnections) are presented through a software interface.. UNIFY develops a set of orchestration engines and controllers that further refine the service graph towards policies and configuration parameters and eventually deploy the resulting NF-FG (Network Function Forwarding Graph) on the virtual infrastructure of the production environment.

Due to the software nature of the virtualized network functions, modern agile software development and operations methods (collectively referred in the industry by the term DevOps), common to software companies such as Google, Facebook, IBM, HP and Yahoo, constitute a good source of inspiration for novel concepts that may be adapted to telecom carrier environments. DevOps relies on four major underlying principles: *Monitor and validate operational quality; Develop and test against production-like systems; Deploy with repeatable, reliable processes; Amplify feedback loops*. Technical aspects associated to these principles reflect on the tools and processes for monitoring, validating and testing software and programmable infrastructure.. Such technical aspects are the focus of our own Service Provider DevOps concept. DevOps has also a cultural dimension, reflected mainly in the Amplify feedback loops principle, which we will not be able to address within this project.

The SDN and cloud management areas in general are hot research topics. We reviewed the literature in line with the areas outlined in the Work Package objectives from the description of work: observability and monitoring; troubleshooting infrastructure problems; verification and network policy checking; testing and debugging of programmable networks. A major problem related to frequent and fine-grained observability updates from many nodes, as envisioned in UNIFY, is scalability and resource-efficiency. At the network level, the applicability of the existing, centralized, Openflow verification tools is limited to the network control plane only. This is a severe restriction in a UNIFY production environment that supports deploying active network functions such as load balancers, firewalls, etc. as part of a service graph. State-of-the-art semi-automated SDN troubleshooting is a mere workflow-led integration of otherwise separated network monitoring or debugging tools. Programming interfaces that enable tools to exchange rich diagnostic data in a controlled manner with components of the UNIFY architecture are required for a higher degree of automation. In addition, virtual network function developers that use one of the recently introduced network programming languages are limited to the network flow space in terms of resources that can be accessed. We identify a need for better support from the infrastructure for controlling and monitoring network, compute and storage resources, as well as infrastructure platform support during the initial deployment and debugging cycles.

We identify four major characteristics of telecommunication networks that make them different from data centres even when a significant part of the functionality is virtualized:

- higher spatial distribution with lower levels of path and equipment redundancy;
- high availability;
- strictly controlled latency;
- larger number of distributed datacentres.

These characteristics pose additional challenges, compared to the state of the art, that need to be accounted for when applying data centre DevOps principles in this environment.

For the Service Provider DevOps concept, we defined two Developer roles: one associated to a classical operator role assembling the service graph for a particular category of services (we call it the *Service Developer*) and another one associated to the classical equipment vendor role in actually programming a virtual network function (we call it the *VNF Developer*). The role of the "operator" in UNIFY is to ensure that a set of performance indicators associated to a service are met when the service is deployed on virtual infrastructure within the domain of a telecom provider. We identify four categories of processes within the WP4 activity areas that involve these actors (in parentheses we indicate the DevOps principle reflected most in a particular process): *Observability (Monitor and validate operational quality)*; *Troubleshooting (Monitor and validate operational quality)*; *Verification (Deploy with repeatable, reliable processes)*; and *VNF Development support (Develop and test against production-like systems)*. To complete the Service Provider DevOps picture, the Bootstrap process from WP2 as well as the Service Invocation and Confirmation processes developed in WP3 need to be involved. Support from the Universal Node is needed for executing the intelligent filtering and aggregation algorithms envisaged by our observability and troubleshooting processes. We detail the WP4-specific process flows by mapping them on the functional architecture defined by WP2.

The processes and their components (exemplified in research challenges and proposed tools) defined requirements and create opportunities for integration between the Work Packages. The sets of initial requirements already reported in the MS3.1 and D5.1 documents are complemented by further description and details in this deliverable. As presented in Annex 2, we conclude that all Work Package objectives, except the one related to the evaluation for which it would be too early in the project timeframe, are covered by several research challenges that will be approached in WP4 as well as requirements placed towards other Work Packages.

We conclude this deliverable by briefly outlining the work towards the next document, Milestone 4.1, planned to be made available by the Work Package in month 12. We need to identify and specify interfaces associated to passing Service Provider DevOps-relevant information between components of the functional architecture. In cooperation with the service instantiation and deployment framework developed in WP3, we will work on specifying how to describe monitoring and verification capabilities such that they could be integrated in the UNIFY production environment. Together with the work on the infrastructure and hardware aspects in WP5, we will work on further understanding how the Universal Node can support our requirements for programmable monitoring capabilities.

# 1 Introduction

## 1.1 Project vision

We envision full network and service virtualization to enable rich and flexible services and operational efficiency. Therefore, the UNIFY consortium will research, develop and evaluate means to orchestrate, verify and observe end-to-end service delivery from home and enterprise networks through aggregation and core networks to data centres. Telecom providers struggle with low service flexibility, increasing complexity and related costs. Although cloud computing and networking have been two active fields of research, there is currently little integration between the vast networking assets and data centres of telecom providers. A unified production environment will create unprecedented opportunities for innovation, an improved quality of experience for users, and technological leadership for European industry and academia. A faster and more flexible network will reduce operating costs and open up new business possibilities.

## 1.2 Relation with other work packages

The WPs of UNIFY and a schematic workflow of the activities are shown in Figure 1. As shown in the workflow WP2 is the main owner of the use cases definition, the related requirements and the architectural aspects so it will integrate and steer the activities of the technical work packages such as WP3 (Service Programming, Orchestration and Optimization), WP4 (Advanced Management Framework and Tools), and WP5 (Universal Node Architecture and Evaluation). In particular, WP3 will work on the definition of service orchestration solution, WP4 will implement a new management framework in the UNIFY architecture, and WP5 in collaboration with the other WPs will design a Universal Node hardware and software architecture and perform an evaluation of its viability. All WPs activities will be highly integrated under the technical supervision of WP2.

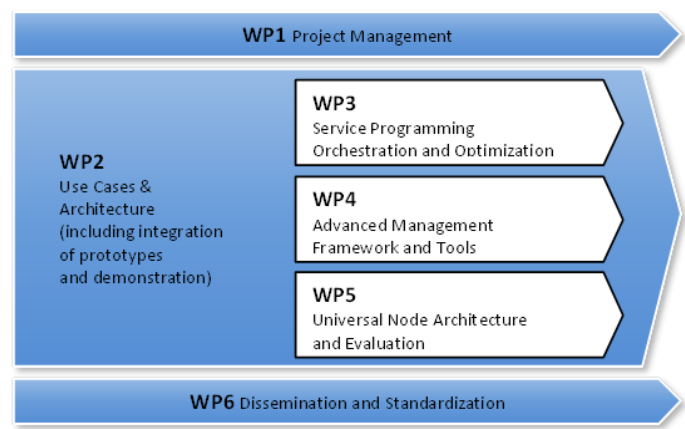

*Figure 1: Relation of UNIFY work packages*

## 1.3 Scope of the deliverable

This deliverable will document the results and status of WP4 at the end of Task 4.1 (Initial Requirements and Specification for Service Provider DevOps (SP-DevOps). In this deliverable, we document the studied state-of-the-art work (Section 2), which includes existing management approaches and models, a description of current service-chain (and middlebox) operation practices with an analysis of the major pain- and cost-points, and a study of related work from the scientific and academic environment related to SDN and cloud management. We will also summarize the status of the UNIFY project based on the current snapshot of deliverable D2.1, describing a major use-case and the preliminary architecture which will enable a carrier-grade environment with high- velocity of feature deployments (Section 3). The deliverable describes a sketch of the SP-DevOps concept (Section 4) developed by applying datacentre-originated DevOps principles to programmable telecommunications networks. Besides the theoretical sketch of how DevOps could be applied in a Service Provider scenario, this section will also describe the identified process embedded into the functional architecture draft of D2.1, and finally give an outlook on specific research questions identified and tools proposed regarding the SP-DevOps processes in focus. A major result of this deliverable is the initial set of requirements (Section 5): technical requirements related to features in the controller and the Universal Nodes, and operational requirements related to interactions between service chain development and operation teams in a telecom provider environment. This deliverable will set the stage for activities in tasks 4.2 and 4.3 and will also have relevance for related tasks in WP3 and WP5.

# 2 State of the Art and related work

In this section we first describe the relevant state of the art in terms of management and operational practises in the industry as background for the UNIFY SP-DevOps concept. We will then give examples of current practices in service operations at telecom providers in order to highlight the major current pain-points. Finally, we will outline the most important related work with respect to management of SDN and cloud. The related work overview will focus on observability and monitoring, verification, troubleshooting, testing and debugging, in line with the plans and vision for the UNIFY SP-DevOps concept.

## 2.1 Management approaches

Management of telecom networks and services incurs high operating expenses related to complex management requirements. However, despite the high expenses, a survey by the TMForum [1] points out that more than 80% of the participating telecom operators indicated that operational excellence is the most important area for improvements. This need for operational excellence together with the upcoming demand for high velocity in service deployment calls for novel concepts, which evolve beyond existing workflows and processes as defined in current frameworks based on historical best-practices (e.g. eTOM and ITIL).

As part of the UNIFY vision, networking and service functions will be virtualized on commodity hardware, and thus treated as software applications. As such, modern software development methods constitute a good source of inspiration for novel concepts. In contrast to network operators, IT companies are already advancing with methods for continuous delivery (CD) today. They are engaged in a continuous cycle to develop code, perform initial testing, release for others to test, deploy in an operational environment, and then monitor the operational status at runtime. These CD methodologies are supported by a set of tools that focus on automation and programmability to accomplish standard management tasks mainly in terms of configuration and fault management in the data centre. The term DevOps is used to refer to the combination of CD methods and supporting tools, although there is no consensus in the industry with respect to the finer details of what is included and what is excluded from this.

In the following subsections, we will first provide the background on traditional best practice models for the telecom and IT industries (i.e. eTOM and ITIL, respectively) and relate them to the goals of the UNIFY project (i.e. SP-DevOps). Next, to highlight the contrast, we will outline the more recent, still evolving, concept for agile DevOps-based IT development and operations. These approaches will form the background for our late discussion on SP-DevOps, an attempt to apply modern DevOps principles on traditional telecom operator business.

### 2.1.1 Best practice models in telecom: eTOM

eTOM [2], the enhanced Telecom Operations Map, introduced by TMForum, defines a best practice model for business processes in the telecommunications industry. An overview is presented in D2.1. The part of eTOM most relevant for the purpose of this deliverable belongs to processes within the Operations area.

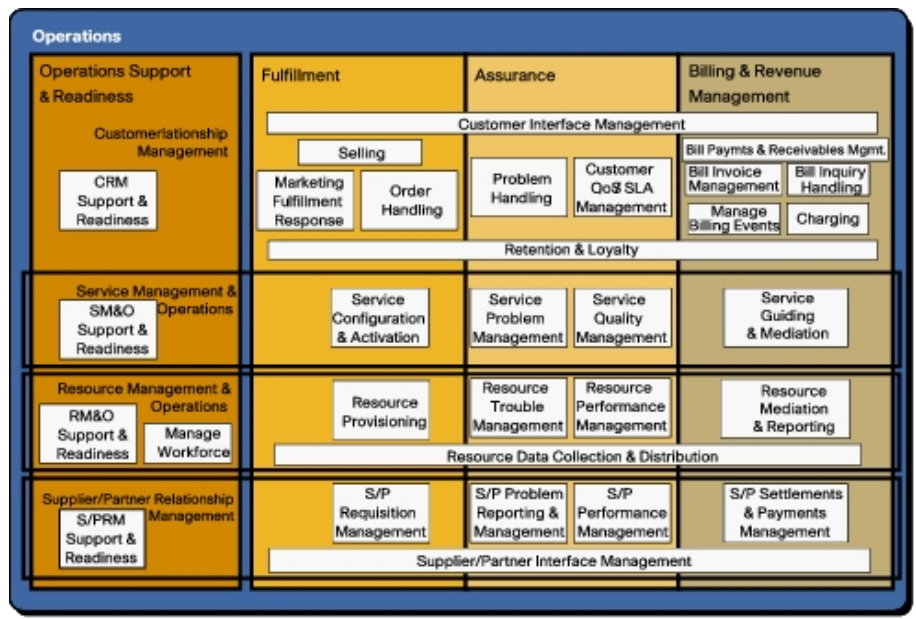

*Figure 2: eTOM Level 2 Model*

Figure 2 depicts the eTOM Level 2 Model for Operations. Here, each core process is generally part of one vertical Level 1 grouping and one horizontal process grouping. This level of detail roughly corresponds to the process descriptions in D2.1 section 6 [3], where the lifecycle of services to be deployed in the production environment is described through a number of UNIFY processes. Most of these processes could be mapped onto eTOM level2 processes in a straight-forward way. As an example, the UNIFY process relating to bootstrapping would map on the vertical process for readiness, covering the horizontal areas of both service and resource management – in eTOM called Service/Resource Management Support & Readiness. On the other hand, UNIFY programmability processes related to orchestration, instantiation and deployment of services dealt within the development of the programmability framework, are mapping nicely onto the Fulfilment processes of eTOM, specifically called Service Configuration & Activation as well as Resource Provisioning. For SP-DevOps, the most relevant UNIFY processes are related to observability and monitoring, troubleshooting, and verification. In the eTOM model, these are encapsulated within the Assurance processes related to services and resources – i.e. Problem Management, Quality Management, Trouble Management, and Performance Management.

From these examples, it can be seen that many UNIFY processes can be mapped onto eTOM – hence the framework proves helpful in order to structure and organize the UNFIY service lifecycle based on industry best-practices. However, from a WP4 SP-DevOps perspective, it is important to highlight the two following observations:

- Operational eTOM processes are defined in the traditional way of encapsulating functionalities within organizational silos, usually pursuing their own goals, with customized tools and an own mind-set (or culture).

This is in stark contrast to the DevOps ideas of cross-functional teams and common tools and goals across the organization.

- While eTOM provides a decent set of best-practice processes for operations of telecommunication services (the Ops part of DevOps), it does not provide the processes for actual development of services and network functions (i.e. the Dev parts). In 2014, TM Forum initiated the ZOOM (Zero-Touch Orchestration, Operations and Management) to address the transition of operations towards DevOps. At the time of writing this deliverable, a set of user stories were made available to TM Forum member companies in an exploratory TR229 report [4].

### 2.1.2 Best practice models in IT: ITIL

ITIL (IT Information Library) v.3.0 is a collection of best practices and guidelines for companies and practitioners on the subject of managing IT services throughout their lifecycle. The applicability of ITIL to telecommunication services is recognized by the TMForum organization and described as part of the FrameworkX [5] . A generic observation is that ITIL describes a series of roles and processes that are considered complex and rigid by the DevOps community. However, we point out that such processes could form a strong baseline for automated actions, while communication barriers between the different roles could be lowered through the use of common tools. The ITIL publications of major importance in the WP4 context are the Service Transition and Service Operation.

ITIL Service Transition defines best practices related to "introducing new and changed services in supported environments" [6]. Of particular WP4-interest are the guidelines for change management, service validation and testing. Release and deployment management could be considered as being addressed partly in the context of WP3. On the other hand, ITIL Service Operation [7] focuses on "achieving effectiveness and efficiency in the delivery of services to ensure value for customer, users and provider". Methods and tools usable for both proactive and reactive operation are provided. Topics of high importance from a SP-DevOps perspective in WP4 are event and incident management, while for example availability of services and optimizing the capacity utilization are considered more of a focus for instantiation and deployment of services dealt with in WP3. We make the following observations regarding the applicability of ITIL to a UNIFY production environment:

- ITIL evolved out of best practices optimized for a manually-driven rigid change management process involving a large number of human actors. In contrast, a DevOps environment is expected to be highly agile and adaptable with focus on small teams.

- While automation is regarded as an optimization option in ITIL, a rather high degree of it is expected in a DevOps environment due to the natural reliance on scripting and APIs. Furthermore, by reducing the number of actors and roles, DevOps methods create an opportunity to further simplify ITIL and thus provide additional gains compared to simple automation.

### 2.1.3 Modern agile development and operations models in IT: CD and DevOps

DevOps is a paradigm shift in the way of developing and operating software and systems, based on close ties between Dev (writing and testing code) and Ops (operating the virtual infrastructure and the application) activities. It appeared initially in startup companies that had to bootstrap with very little human and financial resources. As DevOps ideas spread, a few attempts have been made by practitioners to sketch maturity models (MM) and approaches that cover the gaps between those two activities and show enterprises an evolutionary path. MMs are wide developmental frameworks that enable assessing processes and methods within an organization against a set of benchmark criteria. However, there exists no standardized or universally agreed DevOps model to date. In the following, we will summarize our findings of the most relevant existing MMs and approaches.

For this study, we first have evaluated several proposed maturity models on DevOps and CD. The first three models (i.e. [8] [9] [10]) are relatively similar. They are based on the main phases of the software development and delivery pipeline. Each model contains a subset of {architecture and design, building, deployment, testing, release management, monitoring and reporting, data management} but none of them covers the full set. The model proposed in [10] adds culture and organization as a key area. This is totally sensible as DevOps is at least as much about the mindset and culture around the software delivery process as it is about processes and tools. The model proposed in [11] takes a different approach. It identifies three areas of maturity which do not relate directly to phases of the development/delivery pipeline. Those three areas are Process, Automation and Collaboration. For each of those areas, activities or processes that should be realized are identified.

In all of the above models, five maturity levels have been defined for each area. Two models use generic names to identify the levels (beginner, advanced …), while others use names which are more specific to the achievements that have to be developed to reach that level (repeatable, measured, optimized). It is understood that higher levels in the maturity models should not necessarily be the ultimate objectives to reach by all organizations, as this could introduce unjustifiable costs for the actual benefits. The targeted level depends on the capability, the type of service and application offered and the established business models. For instance, UrbanCode [8] suggested the targeted level for each area and indicated which level is currently the norm in the industry.

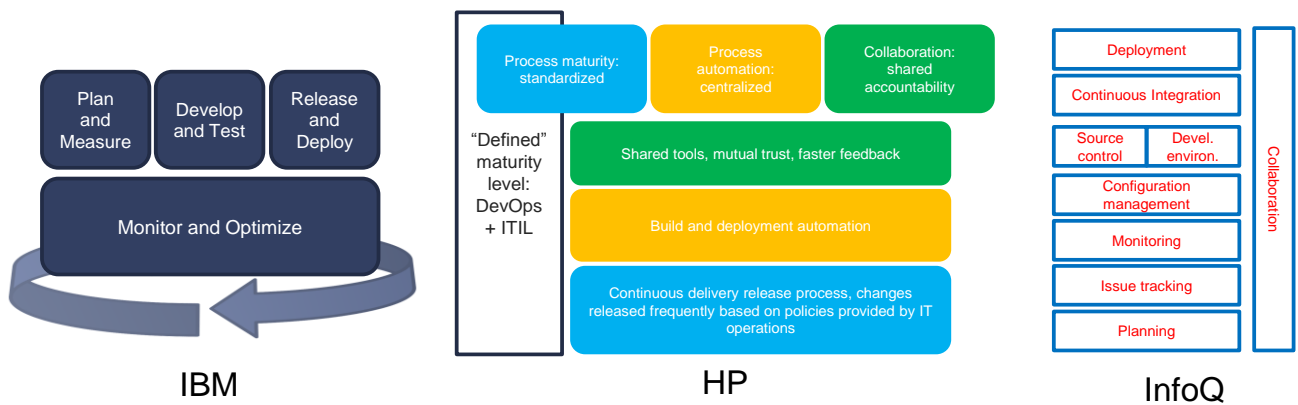

*Figure 3: Comparison of DevOps models*

Besides maturity models, we also considered the high-level descriptions of DevOps by prominent consulting houses. In Figure 3, we compare representations of the DevOps approaches detailed by IBM in [12], HP in [11] and InfoQ in [13]. We observe that the four principles defined by Sharma in [14] permeate through all these models:

- Monitor and validate operational quality
- Develop and test against production-like systems
- Deploy with repeatable, reliable processes
- Amplify feedback loops

As these four principles seem to provide a common ground for understanding DevOps, we will relate to them in our definition of DevOps in a UNIFY environment in section 4.

In addition to the processes outlined by the maturity models, DevOps practices are supported by a range of tools that are used by people in both the development and operations teams, or in mixed teams that include people with dual roles. Popular DevOps tools such as Chef [15], Puppet [16] and Ansible [17] address mainly the configuration management space. Originally restricted to compute resources, they were extended to support the configuration of network nodes. Common characteristics of these tools include the use of templates defined in a domain-specific language and powerful scripting capabilities that allow a high degree of automation. The monitoring capabilities of Ansible are limited to reading counters exposed by the node. None of them was designed for use as a performance troubleshooting or validation tool and therefore are cumbersome and may only provide extremely limited functionality when employed in such scenarios. A recent trend in the industry is to present performance management solutions that were integrated with an SDN controller as DevOps tools [18] [19] from a marketing perspective.

## 2.2 Current practices in service graph operations

In this section, we will review practices in operating service graphs in telecom networks. As dynamic service graphs are yet to be deployed in production, we consider two examples based on fixed service graphs built on physical hardware and an additional example related to the use of middle boxes.

### 2.2.1 Service examples

We present current practices for two examples services, including configuration and maintenance activities. In particular, we focus on the configuration and monitoring of fixed service graphs built on hardware physical network resources: i) an MPLS VPN service which is a rather popular service offered by ISPs to their business and enterprise customers, and ii) an IPTV service, which is also a highly popular value-added service offered by ISPs to their (mainly residential/consumer) customers. The reason for selecting these two services is the fact that their high popularity implies higher occurrence of configuration and maintenance activities; thus, high probability to impact the operating costs of the ISP that is deploying/providing them.

As observed in the eTOM discussion within Section 2.1.1, current best-practises related to the lifecycle management of a telecommunications services do not cover the development of the network functions themselves. This takes place at the equipment manufacturer. The eTOM Product Lifecycle Management covers the work related to defining a new service (such as a MPLS VPN service) in terms of contractual clauses, parameters to be included in the agreement that reflect the service capacity and quality, accounting and billing, etc. These processes are implemented manually by different departments at the operator and rely on the support of a multitude of software tools.

**Example 1: MPLS VPN service**

MPLS VPN employs Multi-Protocol Label Switching (MPLS) to create and support Virtual Private Networks (VPNs), thus, offering the flexibility to network providers to transport and route several types of network traffic with heterogeneous QoS characteristics using the technology and features of an MPLS backbone.

Figure 4 depicts the network infrastructure involved in an MPLS VPN service to the customer of a network provider, including the customer network (right), the customer edge (CE) router, the provider edge (PE) router, the core MPLS network and the access network of the service provider, and finally, the remote user (left) that communicates with the customer.

Specifically, the CE router is practically the customer-premises equipment (CPE) device to which subscribers in the customer's network connect. The CE router connects to a PE router, which is located at the edge of the MPLS core network of the service provider, by initiating a remote access session to it. Then, a PE router connects to one or more CE routers and has full knowledge of the VPN routers associated with each one of them, while it is not aware of VPN routes associated with CE routers which are not connected to it. Moreover, the MPLS core comprises several routers that do not assign VPN information and do not have any awareness of CE routers; the main focus of the MPLS core routers is on label switching.

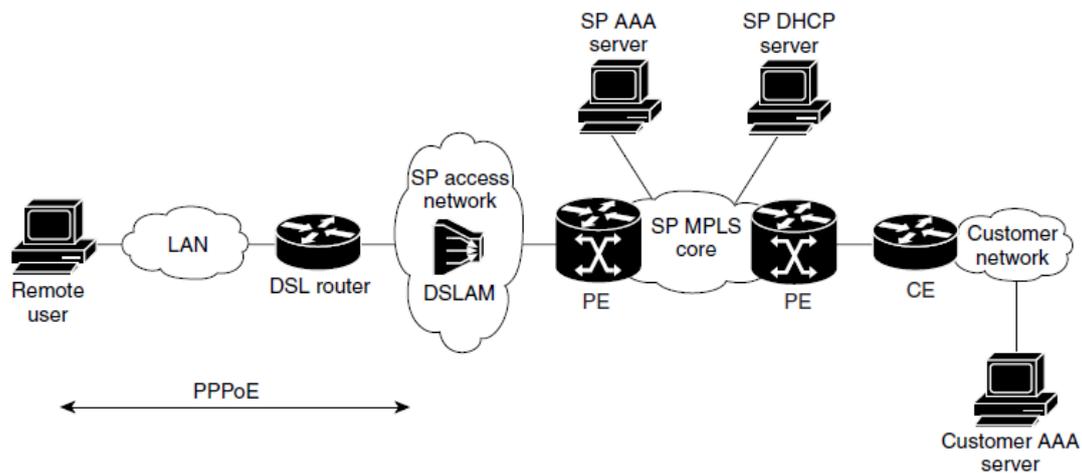

*Figure 4: MPLS VPN network [20]*

The configuration of an MPLS VPN service comprises several steps that need to be completed by a network operator. In particular, the first step is the configuration of the MPLS core network, which includes enabling label switching of IP packets on the interfaces of the core routers, configuration of virtual routes and forwarding tables (i.e. per VPN), association of virtual routes with physical interfaces, and configuration of multi-protocol BGP (MP-BGP) routing session between the PE routers. Limited automation can be implemented based on pre-defined configuration templates, which need to be developed before the first service is deployed and modified to track changes in the infrastructure.

Second, the definition of a virtual template interface which enables the dynamic configuration of virtual access interface (VAI) per user upon request; when the user terminates the session, the VAI goes down and resources are released for other users. Such parameterized templates are defined manually, and applying them to the infrastructure is often a manual task, time consuming and error-prone. Moreover, Asynchronous Transfer Mode (ATM) permanent virtual circuits (PVCs) are created to support encapsulated PPP over ATM on either point-to-point, or to multi-point sub-interfaces. Third, the formation and association of each VPN to a virtual routing and forwarding (VRF) configuration and a virtual template interface is performed. Fourth, the user profiles and Authentication, Authorization and Accounting (AAA) services are configured at the customer premises. Finally, verification is performed according to a manual workbook.

Regarding maintenance, the service provider is obliged to frequently monitor a multitude of interfaces and protocols such as: i) concerning the MPLS core: validation of successful running of the routing protocol, verification of successful label switching, label distribution and bindings, and ii) concerning the MPLS VPN: validation of VRF configurations and routing tables, verification of associations of PE and CE routers, etc. A complete set of tasks related to the configuration, monitoring and maintenance of a MPLS VPN service, in line with the service definition from [20], is provided in Annex 1.

Operational procedures related to a MPLS VPN service are inefficient due to the complex manual configuration (e.g., more than 180 different commands to complete the configuration, even though some of these could be included in templates that are pre-defined) and monitoring of the different nodes associated with the service. Such manual activities are highly time-consuming and prone to errors and mistakes, which in turn imply significant operating costs for the service provider and could be further increased by penalties due to SLA violations w.r.t to delivery or troubleshooting times.

**Example 2: IPTV service**

IPTV has been deployed by multiple network operators for distribution of both live TV as well as video-on-demand (VoD), i.e. on-demand delivery of video content. Unlike (residential) broadband Internet service which is provided on a best-effort basis, ensured high quality-of-service (QoS) is critical for an IPTV service. Therefore, the especially high sensitivity of the IPTV application to impairments creates very big network management challenges.

Figure 5 illustrates the network infrastructure involved in IPTV service delivery to end customers including the transport network, the content delivery network, and the access network [21]. The transport network infrastructure consists of high-bandwidth MPLS/IP core and distribution. A series of specific hardware elements need to be planned, deployed and managed specifically for this service – set top boxes and Video Switching Offices, for example. As Quality of Service demands are not possible to fulfil with a best effort infrastructure, a new virtual network needs to be defined and managed through specific VLAN tags deployed in the DSLAM.

The video head-end consists of real-time encoders/decoders for local and national broadcast video channels, VoD libraries for on-demand video services, and video switching equipment for video transport. The VoD servers implement the storage and real-time streaming functionality for on-demand services. The conditional access system (CAS) provides encryption and decryption services, as well as key generation and distribution functionality, for both broadcast and on-demand services.

The middleware ties a number of logical components together into a more comprehensive IPTV/video software system. The middleware implements the user interface for both broadcast and on-demand services. Note that there are several different middleware implementations depending upon existing/proposed OSS architecture.

Billing of content services can be either pre-paid or post-paid. The end user access is xDSL, or FTTx for wireline providers and QAM/coaxial for cable operators.

The set-top box (STB) is the hardware and common software infrastructure component that is used by the on-demand and broadcast clients as well as by the video decryption function and the video decoder. The hardware may also include a hardware-based decoder and decryption subsystem. The STB software typically includes an embedded operating system, and may also include application infrastructure components such as a Web browser.

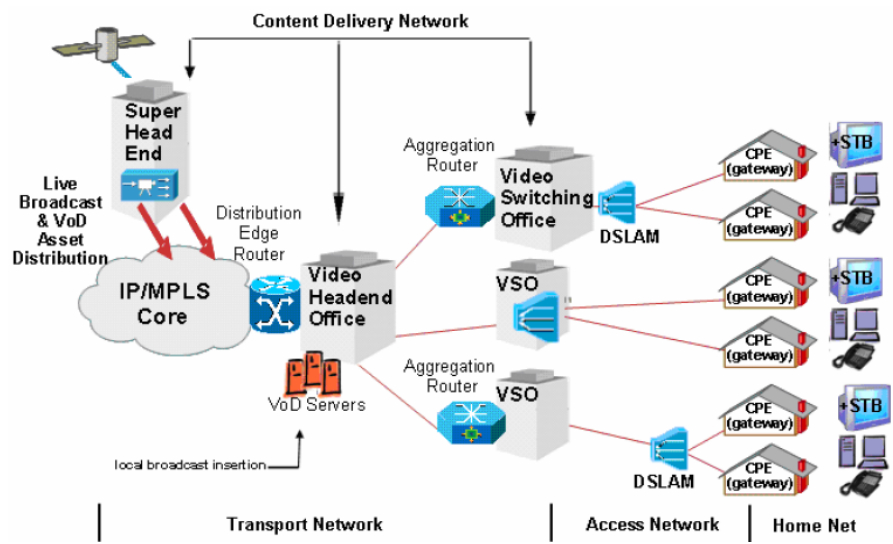

*Figure 5: Network infrastructure involved in IPTV service delivery to end customers [21].*

The service provisioning of IPTV comprises several tasks that need to be fulfilled by a network operator. In particular, the first step is the service activation in multiple informational systems of the service provider such as CRM, DSLAM EMS's, CPE provisioning system, identity and access control system (i.e. AAA), and the billing system. Next, a quite complex configuration must be performed carefully in the head-end, VoD servers, CAS, middleware and STBs. Additionally, proactive and reactive maintenance procedures, fault management and troubleshooting need to be executed so as to ensure that the IPTV service performs according to the QoS levels defined in related SLAs.

Furthermore, IPTV network management puts currently several challenges to service providers. For instance, a service provider must handle multi-vendor equipment, e.g., head-end, middleboxes, VoD servers, CAS/DRM equipment and STBs, which may not be always interoperable. As already discussed, careful configuration must be performed so as to avoid post-installation issues and malfunctions. Moreover, monitoring of the service capacity from the head-end to the access network and instant switching to an alternative (back-up) path are critical to assure high (or adequate) QoS to end-users. Finally, trouble-shooting and isolation of problems is rather difficult due to the complexity of the IPTV system.

Especially, concerning the network management and maintenance, the service provider should perform monitoring of both devices, e.g. the VoD servers, and the service itself, e.g. video quality. Thus, the service provider must monitor a multitude of KPIs such as packet loss, latency and channel change time for the IPTV service, CPU, memory and buffer utilization for the various involved devices, committed information rate (CIR) utilization, as well as queue drops and number of dropped frames for the network. Network management and maintenance is performed by means of mainly two methodologies: i) packet probing, and ii) device instrumentation; both of them are executed manually, thus incurring a high risk for mistakes, or faults to be disregarded.

In conclusion, the complex and mostly manual deployment and configuration, the necessity to involve operations personnel in constant monitoring activities and the difficulty of maintenance, all worsened by the integration complexity of a multi-vendor environment result in time-consuming and very error-prone provisioning of IPTV services, which in turn would implies significant operating costs for the service provider.

### 2.2.2 The effect of middle boxes on management

In current telecommunication networks there is a large deployment of middle-boxes, providing L4-L7 networks services. Current practices are showing that these middle-boxes are implying high capital and operating expenses, complex management requirements, and causes of failures from physical infrastructure and overload. UNIFY goals are to develop an architecture and SP-DevOps solutions in order to reduce capital and operating expenses, improve performance and, maybe, add new types of network services.

Today there is a range of deployed middle-boxes [22] such as WAN optimizers, NAT, proxies, intrusion detection and prevention systems, any sort of firewalls and other application-specific gateways. Each middle-box (typically closed and quite expensive) supports a narrow specialized function (layer 4 or higher) and it is mostly built on a specific hardware platform. Middle-boxes are deployed along most paths from sources to destinations: that's why the Internet lost its initial simple end-to-end forwarding principle. A recent study [23] shows that about 33% of paths tested keep state and perform some level of L4+ functionality.

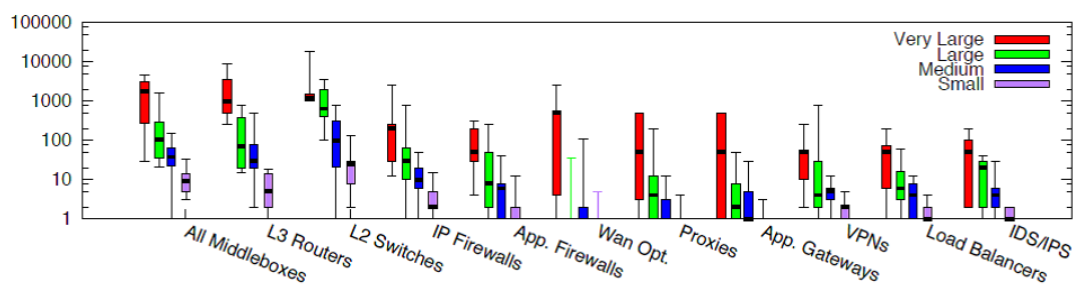

*Figure 6: Box plot of middlebox deployments for small (fewer than 1k hosts), medium (1k-10k hosts), large (10k-100k hosts), and very large (more than 100k hosts) enterprise networks [24]*

In Figure 6 (extracted from [24]) it is shown that the number of middle-boxes is on par with the number of routers in a network. This highlights that middle-boxes today contribute to the network ossification, but also represent a significant fraction of the network capital and operational expenses. The cost is further increased by complex management requirements, and the need for overprovisioning to react to failure and overload scenarios [24]).

For example Figure 7 shows the five year expenditures on middle-box hardware against the number of actively deployed middle-boxes in the network. Figure 8 correlates the number of middle-boxes against the number of networking personnel.

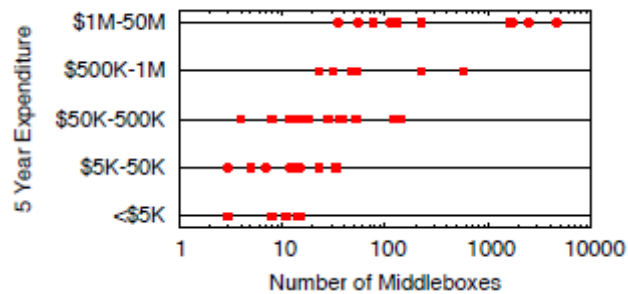

*Figure 7: Administrator-estimated spending on middle-box hardware per network [24].*

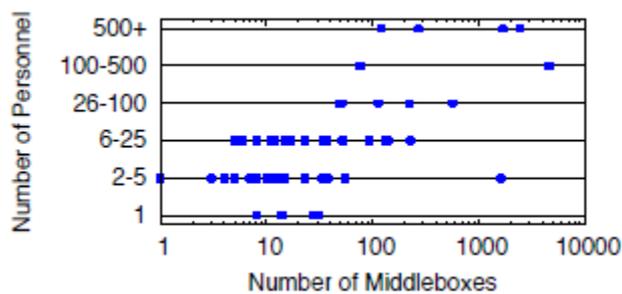

*Figure 8: Administrator-estimated number of personnel per network [24].*

This short analysis illustrates the need for significantly more sustainable and cost-efficient alternatives to expensive and rigid middle-box solutions. The work in UNIFY is aimed at developing a framework based on hardware resources either in the Cloud or in the Network, and with a proper orchestration of virtual elements, which altogether aims to enable efficient, flexible, and dynamic service-chaining and provisioning. In general, there are already Open Source software implementations of firewalls [25], load balancers [26], proxies and caches [27], monitoring and measurement [28], intrusion detection [29] [30], and ubiquitous NAT that support virtualized service-chaining approaches – however, while being a promising start towards flexible service deployment, there are some concerns about, for example, the performance of network functions purely developed in software and running on standard hardware resources. In the next section, two examples further illustrate the limitations in current service deployment and maintenance practices and processes.

## 2.3 SDN and cloud management

The following sections cover previous work within cloud and SDN environments grouped according to the different focus areas of the DevOps efforts in UNIFY : observability and monitoring, verification and policy checking, troubleshooting, as well as debugging and testing. In each subsection, relevant cloud management approaches are discussed followed by currently available SDN approaches.

### 2.3.1 Observability and monitoring

In control theory, a system is called observable if we could reconstruct its internal state based on outputs that can be measured seemingly simultaneously. In a real complex system, such as a telecom network composed of tens of thousands of physical nodes on which millions of software processes are executed, continuous experimental access is limited to only some of the variables that describe internal system states. Such a system is thus only partially observable. In UNIFY, we use the term "observability" to collectively refer to methods that attempt to measure or estimate performance metrics or Key Performance or Quality Indicators and based on them determine particular system states in the UNIFY production environment. Examples of metrics include network delay, jitter and packet loss, processor utilization by a particular process, container or virtual machine; parameters include buffer occupancy, number of flows active, number of containers deployed on the same server; system states include link-level forwarding for a particular flow, network-level forwarding according to a particular routing protocol, the stage in the orchestration process reached by a particular NF-FG at a certain point in time during deployment, whether a particular container is being migrated or not, etc.

Increasing the observability in the network and cloud through the means of resource-efficient and scalable monitoring approaches is an enabler for the deployment and operation of service graphs. According to a recent analyst report taken up in the press [31], up to 74% of the network operations personnel surveyed believe that existing visibility on cloud environments is insufficient. The same survey finds 29% of the administrators complaining of the loss of visibility in SDN environments, 24% find SDN too difficult to troubleshoot and 47% have difficulties keeping up with constant changes. .

#### 2.3.1.1 Cloud monitoring

Monitoring of resources (both virtualized and physical) in cloud computing environments is a mature research area, with production-grade environments operating in large public and private clouds for several years. In this subsection, we review toolsets and framework components representative for public and private cloud monitoring and the monitoring of SDN.

The **JCatascopia** [32] framework includes a series of intelligent probes deployed in the cloud computing infrastructure that store and retrieve the value of an observed metric, a timeline for those observations as well as the capability to perform filtering and adaptive sampling on the collected data. Application-level metrics are supported through a system of plugins that currently support the collection of throughput and delay values. Metrics are aggregated at different levels (compute node or virtual machine levels) and used for further processing via

Agent and Server entities. The framework allows for users to define rules for how metrics should be aggregated. Features supported by the JCatascopia tool are similar to objectives assumed by WP4. The filtering and adaptive sampling capabilities of the intelligent probes, for example, are in line with features we plan to develop at the network level.

**CloudWatch** [33] monitors resources at Amazon, currently the largest provider of public cloud infrastructure. CloudWatch is made available in an "aaS" manner. It includes a large number of pre-defined metrics at both application (for example, MapReduce database transaction statistics) and infrastructure (for example, CPU and bandwidth utilization for a VM instance). The users have a limited capability to define their own metrics, and billing depends on the frequency of the monitoring interval. CloudWatch is integrated with services of the Amazon IaaS platform, of particular relevance being the integration with the Auto-Scaling and Elastic Load Balancing services.

**Hyperic** [34] is part of the VMware cloud management suite. It provides performance monitoring for physical and virtual infrastructure, middleware and several enterprise applications (such as Microsoft Exchange and IBM Websphere) using a combination of agentless and classical agent-based monitoring. Significant capabilities are the auto-discovery of key properties for newly-created VMs and automatic configuration of monitoring functionality. Hyperic also has the capability to copy and re-use monitoring configurations and alert policies, speeding up the deployment of monitoring capabilities. Hyperic is part of the vCenter Operations Enterprise framework [35], which allows determining dynamic thresholds based on hourly-observed behaviour of a performance metric in correlation with other metrics.

Note that both **CloudWatch** and **Hyperic** are proprietary and will not be considered for comparison with the monitoring tools we develop in WP4 as the possibility to determine how specific functionality is implemented is highly limited. We nevertheless include them in order to indicate that market demand exist for intelligent monitoring functions that simplify the work of administrators deploying the monitoring functionality and the analysis of the collected measurements.

**Ceilometer** [36] (also known as Openstack Telemetry component), implements infrastructure-level monitoring in cloud environments based on the Openstack platform. It collects monitoring information from compute, network and storage resources in Openstack-managed environments. It expects the resource managers to publish monitoring information through the Oslo messaging bus, but push and pull agents that communicate directly with the resource managers are also supported. Ceilometer has no capabilities of monitoring virtual network functions, which are regarded as applications from its perspective. Ceilometer offers a REST API to access monitoring information once it is stored in the database. Anecdotic evidence suggests that MongoDB, the database recommended by Ceilometer developers, supports about 386 writes per second and 33,360,480 events per day [37]. Ceilometer is integrated with Heat, the orchestration component of Openstack, and provides input data for performance-triggered auto-scaling rules.

**Google cAdvisor** [38] is a container monitoring tool that can be delivered itself as a container. Newly released in May 2014, the functionality is limited to monitoring only a handful of metrics to analyse usage and performance characteristics of Docker containers. It has programmable filtering capabilities for the data that is generated, although the filters need to be defined at the compilation time of the tool. Both raw and processed data is made available through a versioned REST API.

### 2.3.1.2 SDN monitoring

In today's IP-based networks, OAM tools such as Bidirectional Forwarding Detection (BFD) operate largely in a distributed and decentralized manner [39], which is in contrast to Software Defined Networking (SDN) concepts, which operate with a logically centralized control plane. It has been shown that it is possible to integrate existing OAM tools into SDN (specifically OpenFlow) environments [40]. However, in this approaches the OAM control plane remains distributed which violates SDN principles and complicates management and operations tasks. Furthermore, approaches that rely on pre-Openflow OAM tools require integration of several technology-specific toolsets, which on the long run is not an extensible solution and will substantially increase the complexity of datapath elements. Performance monitoring tools addressing OpenFlow networks were proposed, for example in [4-6]. These solutions utilize the centralized control plane and are implemented in the form of controller applications which take advantage of the existing OpenFlow counter facilities. However, most of these solutions operate under assumptions that are usually not met in a service provider scenario as envisioned in UNIFY: they assume re-active flow instantiation, triggered by the arrival of an unknown flow at a switch; and/or they assume access to fine-grained flow definitions with fully specified matching structures.

**Resource/Accuracy tradeoffs in SDNs** is a topic that is highly relevant in UNIFY to achieve the scalability objectives. Moshref et al [41] explore the trade-offs in resource usage and measurement accuracy for three different SDN measurement primitives: counting, hashing and programming. One main difference between the three approaches is the amount of processing and analysis that is done locally in the switches. In terms of counting, the switches only update flow-based counters and rely on the controller to perform all analysis on these counters and to periodically adjust the measurement rules. A hash-based switch can extract summaries of traffic and transfer the results to the controller for further analysis. A programmable switch can run simple measurement programs to collect and analyse more data locally. The focus in [41] is on a use case where the task is to detect large flows (hierarchical heavy hitters) in a network. The authors argue that at finer time-scales and with more variability in the traffic, hashing and programming offer better resource/accuracy trade-offs.

**FlowSense** [42] is a passive monitoring approach in which information already existing in OpenFlow control messages is used for monitoring link utilization. Based on the information stored in flow control messages upon initiation and removal of flows, the link utilization can be calculated. Passive monitoring is a scalable way of estimating the behaviour in the network, but the limitations are here related to timing - all estimates of the link utilization requires data based on completed flow sessions in order to get all data, which may not be suitable in a UNIFY environment where timing is essential for dynamic service-chains.

**OpenSketch** [43] is an SDN traffic measurement architecture that addresses the challenge of balancing between generality and efficiency in SDN monitoring. The framework introduces a measurement data plane, which is automatically configured by the controller and comprises a three-stage pipeline of hashing (data reduction), filtering (rule-based flow selection), and counting (statistics accumulation). This approach offers a high degree of automation, but the maintenance of 'sketches', i.e., data structures storing information about packet states, takes time and resources. Hence, it might be more suitable for analysis at higher levels in an SDN architecture.

**NetFlow** [44] **and sFlow** [45] are tools developed for flow monitoring in classic IP and Ethernet networks, but could also be applied in cloud and SDN environments. NetFlow-enabled routers collect statistics on IP-traffic data, which are sent to a server for analysis. Such data include e.g. source and destination IP address, ports, type of service, packet and byte counts, timestamps, protocol flags, and routing information. sFlow is a network monitoring protocol that uses random sampling of packets (matching the headers of one or several flows) and scheduled sampling of counters. The results of the sampling are also sent to a server for analysis. One of the differences between the tools is that NetFlow is designed only for monitoring on IP-level, whereas sFlow can be applied in any network layer. As noted in Yu et al. [43], Netflow [44] and sFlow [45] provide generic support for measurements, but the encapsulation and forwarding of datagrams to a controller quickly become too costly in terms of resources in a highly dynamic network environment.

**Latency monitoring** has always been crucial in the operation of a network, and therefore it has been in permanent focus of research. However, SDN creates new challenges (because, for example, the OpenFlow standards do not require too much monitoring capabilities from a standard compliant OpenFlow switch), but SDN also allows implementing some novel monitoring techniques. For instance, in [46] a mechanism to measure link latency from an OpenFlow controller is proposed based on sending a small, specially crafted OpenFlow packet through a link from the controller and back while measuring the amount of time it took to do so. The evaluation shows that their proposed scheme has accuracy close to that of ping but with a lower overhead. In WP4 alternative, scalable and resource efficient approaches will be considered that exploit the features of the UNIFY architecture by either monitoring existing traffic in the network or generating monitoring traffic in the Universal Node (see Section 4.3.1).

#### 2.3.1.3 Conclusion

Monitoring solutions for network and compute resources are inefficient from the point of view of resource utilization and make it difficult or impossible to control trade-offs between accuracy and the resource utilization. In SDN/Openflow, they are limited to using data provided by basic counters supported by the switch specification. Aggregation and filtering algorithms employed are fairly basic and limited with respect to the types of metrics on which they could be applied. Programmability is also a problem, as each of the tools exposes its own interface with individual data and operation encoding, which makes very difficult the integration of multiple tools within a system or accessing data from a different receiver such as orchestration software. Certain problems specific to SDN environments limit the observability when aggregated flow descriptors are used or create a high overhead on the controller. Section 4.3.1 will detail how we plan to approach some of the above challenges.

### 2.3.2 Troubleshooting

Troubleshooting encompasses localization and root-cause analysis of detected faults, changes and performance degradations in the observed network behaviour. The applicability of these and other approaches in cloud and SDN contexts will be considered in the work towards troubleshooting support mechanisms in UNIFY. We use the term "troubleshooting" to collectively refer to techniques that correlate and filter information collected from different entities within the UNIFY production environment with the purpose to identify a particular erroneous situation. Examples of entities include virtual and physical switches, hypervisors and servers. The correlation refers to collecting and assembling together data along a set of rules or descriptions associated to the particular erroneous situation to be investigated. Such data could include, for example, the content of the flow tables within a virtual switch and the erroneous situation under investigation could be the non-forwarding of a particular traffic flow.

#### 2.3.2.1 Cloud troubleshooting

**VScope** [47] is a flexible, agile monitoring and analysis system for troubleshooting real-time multi-tier applications. It allows for dynamically created processing overlays in combination with monitoring and on-line processing of observed metrics. VScope abstracts troubleshooting as a process involving repeated operations such as monitoring of metrics on a set of nodes; interaction between a set of nodes within specified spatio-temporal scope; and analysis of collected metrics from a set of nodes.

**Monalytics** [48] is based on a similar approach as VScope, capable of dynamically constructing Distributed Computation Graphs (DCGs) overlays, implementing monitoring functions for capturing, aggregating and incrementally analysing data on-demand and in real-time. The proposed architecture offers a flexible approach to multi-layer monitoring and troubleshooting at a large scale. Analytic functions can be dynamically created, initiated, adjusted and terminated as necessary and deployed in different types of centralized, hybrid and tree topologies for meeting different requirements on costs and analytic needs.

**MonitorRank** [49] provides a ranked order list of possible root causes of detected anomalies in a service-oriented web architecture. The aim is to isolate and rank the combinations of services and API calls that are most likely to be a root cause, without complete prior information about dependencies or domain knowledge. For this reason, the approach relies on unsupervised learning and is partially based on pseudo-clustering and analysis of historical time-series of monitored metrics and API calls in addition to generated call graphs. A random-walk approach is employed to compute a score used in the ranking process performed in obtained call graphs.

### 2.3.2.2 SDN troubleshooting

**OF-Rewind** [50] is a SDN-debugging tool that is capable of recording both control and data traffic traces of an OpenFlow network, and is capable of replaying it in a custom OpenFlow network to reproduce the bugs. The challenges in replaying include timing accuracy, multi-instance synchronization, and online replay of multiple network elements. The amount of traffic passing through a network can be significant, and for this reason OF-Rewind records a subset of traffic and uses this to reproduce or to find the root cause of bugs. However, partial recordings can be insufficient as some data needed for replay and debugging may be missing.

**Automatic Test Packet Generation (ATPG)** [51] is another debugging tool, which automatically generates a minimal set of packets in a real network to test all forwarding rules, firewall rules, links and network elements for errors. For finding the minimal set of packets, ATPG takes snapshots of forwarding tables periodically. The efficiency of ATPG approaches depends on the snapshot (or probing) interval and is thereby linked to the additional load incurred in the network. The approach in this case is restricted to only the action part of the forwarding rules and does not account for the matching part, meaning that relevant debugging information may be missed. The approach requires additional fields that are not currently supported by the OpenFlow standard for troubleshooting purposes.

**A systematic troubleshooting** methodology for SDN is suggested and discussed in [52], in which state and code layers are related to corresponding categories of faults. In the proposed layered approach a binary search troubleshooting procedure can be employed to localize a bug (e.g. erroneous control logic). The authors outline a generic troubleshooting workflow that can be implemented and executed automatically, and that can be combined with recently developed troubleshooting tools in each step (such as ATPG [51], OF-Rewind [50], NICE [53]). The overall methodology is in general relevant to consider with respect to the troubleshooting and verification approaches investigated in UNIFY.

**The SDN Troubleshooting System (STS)** [54] automatically reduces the sequence of debugging events (isolated by the use of e.g. OFRewind) and other important events (like link and component failures) to a "minimal causal sequence" that still triggers the same bug. The main contribution is the generalization of delta debugging to distributed systems which enables STS to prune unnecessary events. STS also benefits from internal events exposed by the controller software. Additionally, STS relies on the Hassel library [55] to notice when a bug manifests. A limitation of this off-line tool is that it is inadequate in dealing with performance bugs.

### 2.3.2.3 Conclusion

Troubleshooting in SDN is an active research area, with many problems being identified and point solutions being independently proposed to address them. However, daily operations of resources would thus require expert knowledge of tens or hundreds of tools that each diagnoses small categories of problems. The integration of such tools into workflows is difficult because of the lack of common interfaces. Many of the tools require gathering all data at a centralized location or massive generation of test traffic, which puts a high load on the infrastructure. Existing methods rely on the OpenFlow specification, but visibility on other infrastructure control interfaces need to be considered in a unified production environment. . Section 4.3.3 will detail how we plan to approach some of the above challenges.

### 2.3.3 Verification and policy checking

In UNIFY, we use the term "verification" to collectively refer to approaches that compare and contrast expected and detected system states in the UNIFY production environment. The expectations are based on pre-defined representations of the system states under investigation, while the detection of the actual state can be provided through observability or by other means. Examples of such system states include availability of cloud infrastructure, compliance with security policies, the existence of forwarding loops or whether a certain node is reachable or not. Mechanisms for verification and policy checking are necessary to avoid conflicting behaviour in the network devices caused by contradicting rules (such as forwarding loops) or violations against e.g. security policies. Note that the following approaches in cloud environments are mostly related to application-level services.

#### 2.3.3.1 Verification of cloud computing functionality

**Verification of cloud services** and an overview of existing techniques are provided in [56]. In this paper, the authors investigate existing tools and methods for cloud consumers and service providers to verify that their services work as expected from different points of view: functional correctness, service availability, reliability, performance and security guarantees. Given the narrowness of the existing tools, the authors encourage future efforts on this research area and highlight some promising directions. For example, some work has been done in the direction of verifying the users interaction with untrusted cloud services (Venus [57], SPORC [58]), verifying sensitive data propagation within a cloud environment (CloudFilter [59]) and solve accountability issues that raise when moving services to a cloud infrastructure [60] [61]. However, authors identify different areas where we still lack specific cloud-oriented verification tools and they propose some possible promising research directions to help and encourage future efforts exactly on these issues. To this end, from a functional point of view, it would be interesting for customers to have tools that check whether the cloud infrastructure is operating in the correct way, i.e. if it is running the correct application. In addition, customers need to verify that performance and availability levels are aligned with the ones agreed with the cloud provider by means of SLAs and, if a violation occurs, they need to assess how frequently it happens.

**Verification of multi-domain cloud security policies** is discussed in [62]. In this paper the authors define a model checking technique that can be used as a management service/tool for the verification of multi-domain cloud security policies. The necessity of a collaboration scheme among different cloud systems is becoming more and more important since this enables them to achieve higher uptime and services usage.

**Cloud services composition** is discussed in [63] where a framework for Cloud service composition is introduced that aims to overcome the issues caused by the open and flexible nature of Cloud services, by incorporating some trusted third-party entities to govern and optimize the service composition process.

20          Deliverable D4.1          10.02.2015

### 2.3.3.2 Verification of SDN functionality

**FORTNOX** [64] is a software extension to the NOX OpenFlow controller that aims to enforce the security constraints imposed on an SDN network by a security application. In particular, FORTNOX prevents an application to inappropriately install new rules that contradict the existing ones. To achieve this goal, FORTNOX detects and resolves rules conflicts at execution time by analysing each newly arrived rule. The conflict resolution algorithm can perform on-the-fly checking of hundreds of rules with an overhead in the order of few milliseconds. FORTNOX is targeted at detecting conflicts between security-related rules, but the approach is relevant for verifying other aspects in the deployment of a service-chain in UNIFY.

**VeriFlow** [65] is a layer between a software-defined networking controller and network devices that dynamically checks for network-wide invariant violations at each forwarding rule insertion. In particular, VeriFlow introduces novel incremental algorithms to search for potential violation of key network invariants — for example, availability of a path to the destination, absence of routing loops, access control policies, or isolation between virtual networks. The approach is based on checking various network properties for the packet equivalence classes affected by the new rule, followed by building a forwarding graph representing forwarding decisions at each node, that in turn is validated against the network invariants. Some of the limitations of the approach are related to scalability and accuracy as it is assumed that a complete view of the network is available through a centralized controller which may work well only for smaller networks.

**NetPlumber** [66] is a real-time policy checking tool with the Header Space Analysis (HSA) as its theoretical foundation. It is targeted at verifying SDN networks, but its abstract concepts can be adapted to fit conventional networks as well. Header Space Analysis is based on a geometric interpretation of the packet headers as an L-dimensional space and on switch processing as a transfer function that operates on this space. Applying the switch functions in cascade, the system can explore and analyse a variety of reachability properties. Although NetPlumber has proved to be scalable and efficient (in real network deployments with thousands of rules), the main limitation of NetPlumber (as well as HSA, in general) is that it cannot precisely model dynamic network appliances or network functions, since it relies on reading the state of network devices. An additional limitation could be the large amount of time needed to handle link up/down events, due to the regeneration of the NetPlumber internal network representation.

**Header Space Analysis (HSA)** [55] allows for static checking of the forwarding rules of a whole network. It can detect reachability failures, forwarding loops, traffic isolation and leakage problems. Moreover, other tools (ATPG [51], STS [54], [67]) have already been built upon the open source implementation of the proposed technique called Hassel. The key to HSA is the efficient calculation of Network Transfer Functions, which tells how the bits of incoming packet header are changed when the packet passes a specific forwarding element. Although the Hassel library seems practical for analysing a campus-sized network, HSA is currently not capable of handling nodes with stateful forwarding rules.

#### 2.3.3.3 Conclusion

The majority of the SDN verification tools operate on network configuration rules (commonly OpenFlow), and none of them considers active network functions (i.e. VNFs or middle-boxes that dynamically change the forwarding path of a flow according to local algorithms, as for example an intrusion detection system or a load balancer may do). In other words, these tools operate on the (centralized) programmability of the control plane only and are not adequate for situations such as the UNIFY production environment where VNFs may program the data plane and cloud resources are integrated with the network. Section 4.3.2 will detail how we plan to approach some of the above challenges.

### 2.3.4 Testing and debugging

Deployment of applications and network functions requires various types of debugging and model-checking tools to ensure the intended behaviour of the component. In UNIFY, different approaches to testing and debugging service-chains and included service-components are considered and leveraged from existing methods. From a cloud management perspective, cloud testing is the concept of evaluating the functionality and performance of software and services using a cloud platform, in which different scenarios and traffic conditions can be tested in a large scale. A survey of representative approaches for cloud testing (including multi-layer testing, SLA-based testing, large scale simulations and on-demand test environment) is presented in [68]. The paper highlights that the focus is shifted from software development and product-oriented activities to service-oriented reuse, composition and online renting. The challenges of cloud testing is in terms of the dependability between systems, parallelization, and development of evaluation metrics, to name a few.

#### 2.3.4.1 Cloud testing

**CloudTest** [68] is a production software tool for functional, load, and performance testing of web sites and web applications. Nodes can be distributed across public and private clouds to cooperate in a large load testing. CloudTest On-Demand allows for testing web sites under normal and extreme traffic conditions and can be used to simulate thousands of virtual users visiting website simultaneously, using either private or public cloud infrastructure service. Memory-based techniques enable real-time analysis of the data produced in large-scale tests. Provisioning data are displayed via an analytic dashboard on a synchronized time-line.

**iTKO LISA** [69] provides a cloud-based environment and services for composite application development, verification and validation. The framework builds executable test cases for functional, load, and performance testing, which enables complete tests of the service. The testing capabilities of the framework include coverage-based testing for heterogeneous distributed architecture, codeless testing, UI testing, as well as load and performance testing. LISA also provides a codeless testing environment for QA and development which enables a rapid design and execution of automated tests.

**Cloud9** [70] is a cloud-based testing service that uses parallel symbolic execution techniques by migration to the cloud platform, and implements the "Testing as a Service" (TaaS) concept. Under this model, users of the Cloud9 testing service are charged according to test goal specifications that they provide, meaning that the total cost of service is proportional to the program size. During a test, the Cloud9 attempts to follow all the possible executions by exploring the program path-by-path, with the drawbacks of extensive path exploration, CPU-intensive constraint solving, and high memory usage. Although this is offered as TaaS, the approach is relevant to SP-DevOps as an example where the execution of a software or service component is divided and tracked across several nodes in, for example, a sandboxed testing environment.

**Cloudstone** [71] is an open-source toolkit consisting of a set of automation tools for generating load and measuring the performance of a synthetic Web 2.0 social application in different deployment environments. The toolkit allows for selecting a deployment architecture and automated deployment of components; specification of workload profiles; experimentation using a workload generator; and analysis of the resulting behaviour. A metric in terms of dollars per user per month is introduced for evaluating the cloud performance in terms of usage and costs relative to e.g. consumed storage and computing. These approaches are focused on testing software developed for cloud computing but include similar challenges related to testing of service chains and individual components within the DevOps-concept. The challenges are similar in terms of virtual machine placement, dependency relations, and traffic generation as well as testing in a secure environment that will not affect other parts of the cloud system.

### 2.3.4.2 SDN debugging

**NICE** [53] is a model checking tool, augmented with a symbolic execution engine, capable of checking the correctness of an OpenFlow application, based on the popular NOX controller, at development time. NICE can test an unmodified version of the controller application using a simplified model of hosts and switches against a number of predefined properties, possibly extending them to meet the user's requirements. The heuristic approach employed in NICE makes the method fast and efficient in finding several types of bugs, such as routing loops and black holes, and it can also verify different reachability constraints. The main limitation of NICE is related to the time duration of a complete verification process in a real (and complex) network scenario and to the memory requirements of this process, which will likely not meet the requirements of the UNIFY use cases with respect to real-time verification in fast DEV-OP cycles.

**OFTEN** [72] test switches and controllers together as one system, instead of testing individual components of an SDN system. The authors argue that, for example, in case of OpenFlow switches, the compliance validations to OpenFlow standards are simply not enough because of cross-feature interactions among different components of the system. Traditional testing of an integrated network is more or less straightforward, but it is a tedious task that includes writing lots of specific test cases. OFTEN takes another approach, it modifies NICE by extending its controller testing capabilities to the whole system. The model checker of NICE uses a model of the network environment, whereas OFTEN interacts with the real network. One limitation is that it is not possible to know for sure whether an OpenFlow switch has finished processing a datapath packet. Preliminary results show some potential of the tool, but the size of state space limits its practical usage in case of complex networking settings.

**Network debugger (NDB)** [73] is a network debugging primitive analogous to well-known software debugging ones. The debugging tool allows users to study packet backtraces when a user-defined breakpoint is triggered. Breakpoint conditions can be based on packet fields (like OpenFlow matching rules) and packet paths; a packet backtrace is a sequence of forwarding actions of the packet. NDB modifies the OpenFlow control traffic in order to duplicate packets in the switch for each matched flow entry. The duplicates are then marked and sent to a collector module, where the unique marks help to create backtraces. Although there are some cases when the correct backtrace cannot be reconstructed due to some ambiguity, SDN developers may still find NDB useful in case of diagnosing bugs in the controller or in the switch that affect the correctness of forwarding.

**Anteater** is a debugging tool [74], which takes snapshots of forwarding tables of distributed or centralized networks and analyses them for errors. Anteater converts forwarding table states into instances of Boolean satisfiability (SAT) problem, and uses a SAT solver to find bugs. Anteater can verify a network for the reachability, forwarding loops, and packet loss issues without sending real packets in the network. Although the detection performance is rather high (86% of randomly sampled bugs from the Bugzilla open-source repository), one of the limitations of Anteater is that it assumes that the snapshots in a network are consistent over time. However, in large networks with frequent changes to a forwarding state, the snapshot might be inconsistent because the network state changes while the snapshot is being taken. The other limitation of this tool is that it cannot test issues due to link failures or hardware failures.

### 2.3.4.3 Conclusion

Testing and debugging tools aimed at clouds and SDNs have a series of limitations in terms of the explosion of states that need to be examined, the assumptions they make regarding the stability of the network and the scalability of the probing approach. No environment that tests both cloud and SDN resources exists at the writing of this deliverable. Section 4.3.2 will detail how we plan to approach some of the above challenges.

### 2.3.5 Distributed SDN control planes

Openflow started as a logically and physically centralized architecture and evolved towards different levels of logical centralization and physical distribution. For carrier networks as complex and large-scale as those envisioned to be supported through the UNIFY architecture, it is inevitable to distribute the network control plane across several compute and storage resources, which can (and should) be spread geographically as well. This section will introduce some examples on how distribution of functionality has impact on observability, verification and troubleshooting capabilities of the network.

In [75] the authors identify two main approaches to distributed SDN control planes; *flat* or *hierarchical* (Figure 9). The authors identify classes of control plane functionality which do not rely on a global view of the network, and thus could be handled by a local controller, i.e., a controller which is topologically close to the network devices it controls. One example of control plane functionality which can be handled locally is heavy-hitter detection, an important feature for network troubleshooting in WP4, intrusion detection, traffic engineering and load-balancing.

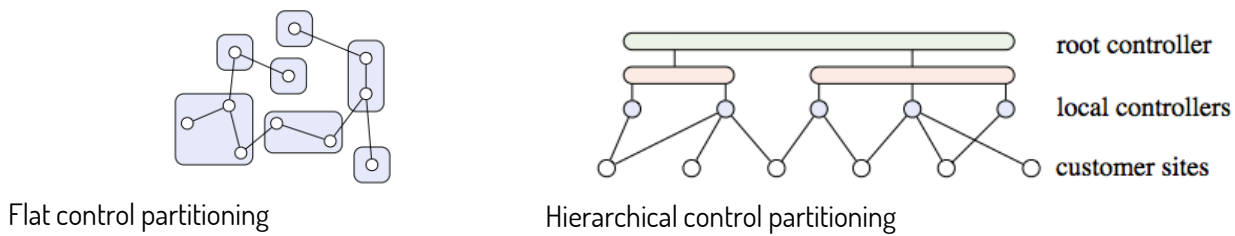

*Figure 9: Approaches for distributed SDN control planes*

Flat SDN control is required to realize a cooperative control plane across administrative, topologically, or geographically diverse datacentre locations. Flat control plane partitioning allows for global optimizations, such as optimized placement of content caches or other chains of VNFs. The amount of state required to synchronize across all partitions however can be kept low by limiting it to state which is required for global optimizations. Determining what state needs to be shared across a flat control plane to facilitate troubleshooting of faults that affect multiple domains is relevant in the WP4 context.

A hierarchical distribution of a WiFi-tailored SDN control plane was presented in [76], providing fine-grained control over rapidly changing transmission parameters that are specific to the IEEE 802.11 WiFi protocol. Another example is the ElastiCon [77] approach which is an elastic distributed controller architecture, which addresses the issues that may arise due to the static configuration of the mapping between switches and controllers, i.e., uneven load distribution and lack of elasticity. Approaches intended for operation in decentralized architectures are also considered, addressing for example non-conflicting policy composition in distributed control planes [78].

From a WP4 perspective, the decentralized and distributed operation is highly relevant and necessary with respect to scalable and resource-efficient network observability and troubleshooting. This is achieved through a hierarchical controller architecture operating based on different layers as described in WP2, but also by enabling distributed node-local analytics in the Universal Nodes developed by WP5. Verification and debugging could be performed in a centralized and decentralized manner, but may require different assumptions and approaches, which are subject to further investigation.

# 3 Summary of relevant UNIFY results

In this section, we will briefly summarize the initial considerations regarding UNIFY use-cases, processes, and architecture aspects[1]. We will first present the use-case selected as the main focus of this work-package, and highlight how it is suitable to highlight relevant WP4 aspects. We will then outline the envisioned UNIFY service lifecycle and related processes selected as the main scope of the project, in order to give a better understanding of how SP-DevOps will fit into UNIFY. Finally, we will provide an overview of the initial overarching architecture together with a first draft of the functional architecture, derived from the selected processes of the UNIFY service lifecycle. This will allow us to describe SP-DevOps process flows embedded into the UNIFY architecture and point out research challenges in Section 4.

## 3.1 Exemplary use-case: Secure, content aware IP VPN

This example shows a potential future evolution of the baseline MPLS VPN case described as state-of-the-art in section 2.2.1 to include additional functions such SSL accelerator, malware and intrusion detection and private content delivery network (Figure 10). If such a complex service would be realized with today's production-grade telecom hardware, it would require that the MPLS VPN infrastructure be supplemented with a number of middleboxes, each middlebox dedicated to supporting one network function. Apart from the disadvantages incurred with managing hardware middleboxes that were mentioned in section 2.2, placing the hardware in fixed locations within the network would require careful planning from the operator and has the potential to introduce additional delays in order to channel traffic from customer premises to the fixed location where a particular hardware box was deployed. The installation and operation of the middleboxes is equivalent to building a parallel network (similar to what had to be done for the IPTV service described in section 2.2.1), dedicated to this service, and having a small potential of reuse for other services.

D2.1 described a Content-Aware IP VPN use case where a carrier provides value to a large enterprise customer by embedding Virtual Network Functions (VNFs) corresponding to SSL accelerator functionality, a malware and intrusion detection function, private content delivery network support and an elastic router (Figure 10). Compared to the rather static middlebox-based MPLS VPN extension, the UNIFY Content-Aware IP VPN example shows how the operator can use Virtual Network Functions deployed on generic hardware instead of building a parallel network infrastructure to support this service. On-demand VNF placement and re-use of generic hardware between different services are facilitated. The UNIFY Content-Aware IP VPN has an inherent dynamic aspect in that policies associated to it allow the enterprise to increase the resource usage 10x for a relatively limited time interval, in the order of minutes or hours, by employing an elastic router and taking advantage of a programmable optical transport network. Traffic that was identified as infected documents by the malware-IDS function is automatically and dynamically made to bypass the content delivery network, thus reducing the spread of an infection.

---

[1] These considerations are taken from UNIFY D2.1 [3], which represents the current status of the integrated input and work carried out in all technical workpackages WP3, WP4 and WP5.

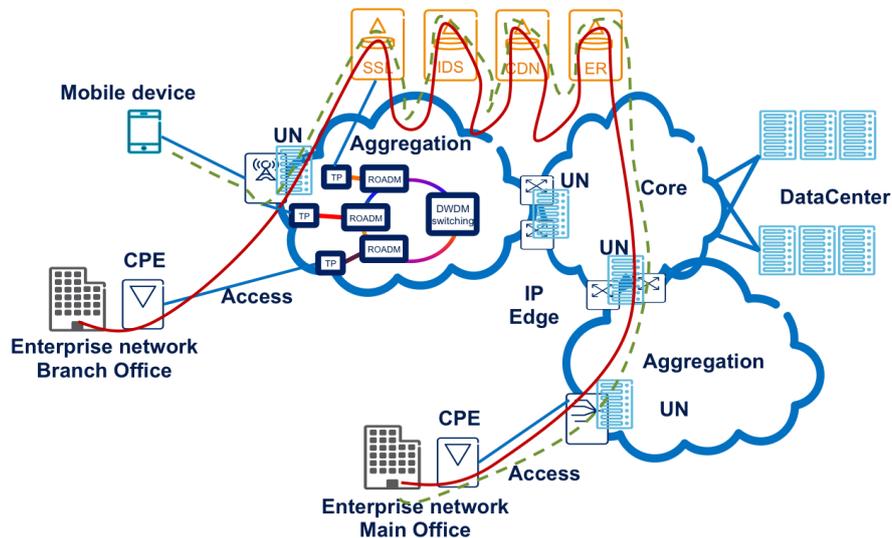

Figure 10: Secure, content aware IP VPN service [3]

A Service Graph is an association of VNFs and their interconnections. Some examples of basic monitoring functionality internal to the Service Graph include compute resource utilization for the SSL VNF, compute and storage resource utilization for the IDS VNF, network delay measured between the service access point at the branch office and the service access point at the main office, packet losses measured along traffic tunnels within the aggregation and core parts of the network, etc. Key Quality Indicators that are agreed through the Service Level Specification are usually exposed to the enterprise customer through a self-service portal. Examples of such parameters could include Committed Information Rate (1000 Mbit/s and 100 flows/s), Committed Burst Rate (10000 Mbit/s, 30000 flows/s for a time interval of 1h), Malware detection rate 98%, Caching maximum capacity 10 TB, branch-main office latency 10ms.

The following aspects of the use case are relevant from a SP-DevOps perspective:

- An elastic router that increases and decreases capacity depending on the traffic demand requires that monitoring functionality associated to it scales to a) include the new instances added to the router and b) accommodate a potential increase in monitoring-related traffic generated by the new router instances (such increase could be significant if certain ports had to be put in span or mirror modes, or traffic is being captured through technologies such as sFlow or IPFIX). Every scale-in and scale-out operation requires that flow descriptors for incoming and outgoing traffic from the various instances are verified as being correctly configured in the infrastructure.

- Software updates (for new functionality or bug fixing) are expected several times a year. Critical security updates needs to be applied outside pre-scheduled maintenance windows. This creates a need for capabilities supporting the testing of such new patches in an environment as close as possible to the real operating network. Such capabilities may include monitoring features that provide a higher level of detail than during normal network operations. They could also include verification features that validate both the deployment and the resource usage in the test environment.

- The fact that traffic dynamically bypasses certain VNF instances (for example, traffic identified as malware-infected documents are not sent through the content delivery network) requires a verification capability that is able to validate such cases.

- The telecom operator has the capability to rapidly customize the service for a particular customer based on diverse criteria, such as using VNFs that have restricted features but are more cost-effective for the customer or choosing VNFs provided by a particular software company in case policy restrictions demand it on security grounds.

## 3.2 UNIFY process model and service lifecycle

The UNIFY service lifecycle has been derived from traditional operator processes (eTOM) as well as DevOps principles in order to accommodate for increased dynamicity and higher service velocity. We introduce the model in Figure 11. A detailed description with the relationship to our developed SP-DevOps concept will be presented in Section 4.

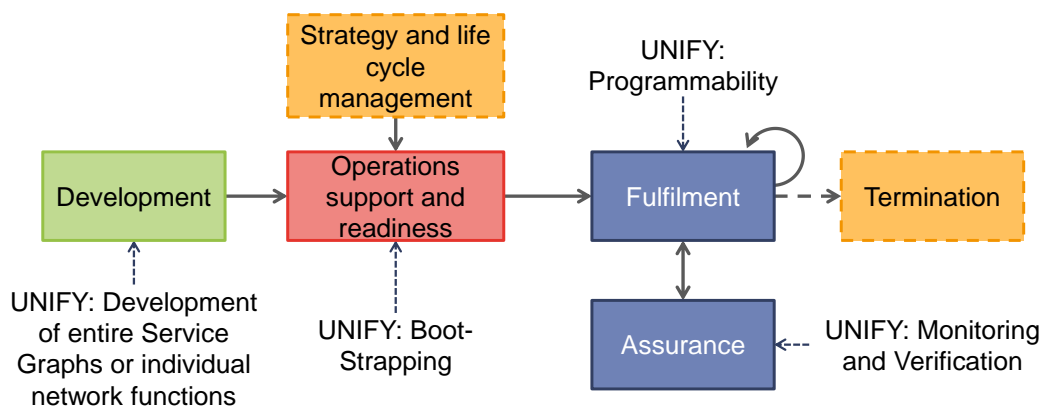

Figure 11: Simplified Process model and mapping to UNIFY

As the main scope of the UNIFY project as a whole, we identify four larger process groups relating to Development, Operation Support and Readiness, Fulfilment, and Assurance. In an eTOM analogy, these processes would be situated on the Level 1 of the framework. The SP-DevOps concept will span the four main UNIFY processes and provide an additional level of detail focused on particular areas. In an eTOM analogy, the SP-DevOps processes would be situated on the Level 2 of the framework. The focus will be on the assurance part providing additional management capabilities to the service and virtual network function graphs in the areas of monitoring and verification. The UNIFY high level processes served as guidance and input for detailing the functional architecture (see Section 3.3) and can include further sub-processes:

- <u>Boot-strapping</u> is covering all aspects to enable the unified production environment to work. This includes details required to translate and map information between architecture layers and abstractions, information about the resources and it's capabilities,, to inform relevant elements on the different architectural layers about available information and to enable infrastructure to perform basic management tasks such as monitoring of nodes, links and interfaces between architecture layers.

- The programmability framework will rely on available information and interfaces from the boot-strapping process. The goal with UNIFY's programmability framework is to enable on-demand processing anywhere in the physically distributed network and clouds. The major objective is to enable dynamic and fine-grained service (re-)provisioning, for which a <u>Service Invocation</u> sub-process will resolve and map network function forwarding graphs (NF-FG) and associated requirements through different levels of abstractions (and virtualizations) to physical resources available in the distributed system (both network and cloud) while adhering to operational policies..

- Monitoring and verification (more details in Section 4) is covering a set of operational aspects for service and virtual network function graphs like collection of information, verification, analytics, failure detection and resolution, quality observations. These processes will rely on information from boot-strapping process, as well as interaction with programmability for monitoring function placement and configuration. Here, sub-processes for <u>Observability</u>, <u>Verification</u> and <u>Troubleshooting</u> have been defined, all focus areas of the work in WP4.

- The Development process is capturing the aspects of the development (or rather definition) of service graphs on the one hand as well as dynamic development of network functions (VNFs) on the other hand. In UNIFY, we assume that the needs of the service developer are largely fulfilled with support of monitoring and verification functions introduced above. The additional requirements of VNF developers will be covered in an additional process specifically aiming at <u>VNF development support</u>.

Additionally, we have identified two other processes in the eTOM process model which are relevant for the completion of the UNIFY service lifecycle, but which have been considered out of scope in terms of research focus within the project: Termination process, and Strategy and live cycle management (shown in Figure 11 as orange boxes with dotted lines). But both processes are considered out of scope in terms of research focus within the project and therefore will be not further detailed.

## 3.3 Initial UNIFY Architecture

In order to understand the SP-DevOps process flows detailed in section 4.2.2, we will briefly introduce the initial draft of the UNIFY architecture. UNIFY is defining a three layers architecture as reported in Figure 12, composed by a

29                                        Deliverable D4.1                                        10.02.2015

Service Layer, an Orchestration Layer and an Infrastructure Layer. The three layered model can be augmented with an application layer corresponding to the users of the services (shown as Service + SLA).

The Service Layer is in charge of turning the service chain provisioning into consumable services by defining and managing service logics; establishing programmability interfaces to users (residential, enterprise, network-network, OTT, etc.) and interacting with traditional OSS/BSS systems. The service layer is also responsible to create further service abstractions as needed toward the different users (e.g., BigSwitch topology) and to realize the necessary adaptations according to such abstractions.

The Orchestration Layer is split into three major sub-components: the Resource Orchestrator Layer, the Controller Adaptation Layer and the Controller Layer. From a resource orchestration point of view, the Orchestration layer collects and harmonizes virtualized resources into a global virtualized resource view at its compute, storage and networking abstraction. The global resource view in the orchestrator consists of four main components; forwarding elements, compute host capabilities, hardware-based or accelerated network function capabilities, and the data plane links that connect them. All of the resources must have some associated abstract attributes (capabilities) for the resource provisioning to work. The Controller Layer is responsible to provide technology independent control interfaces and to virtualize resources. Note that there could be as many different controllers in the Controller Layer as there are different technical sub-domains.

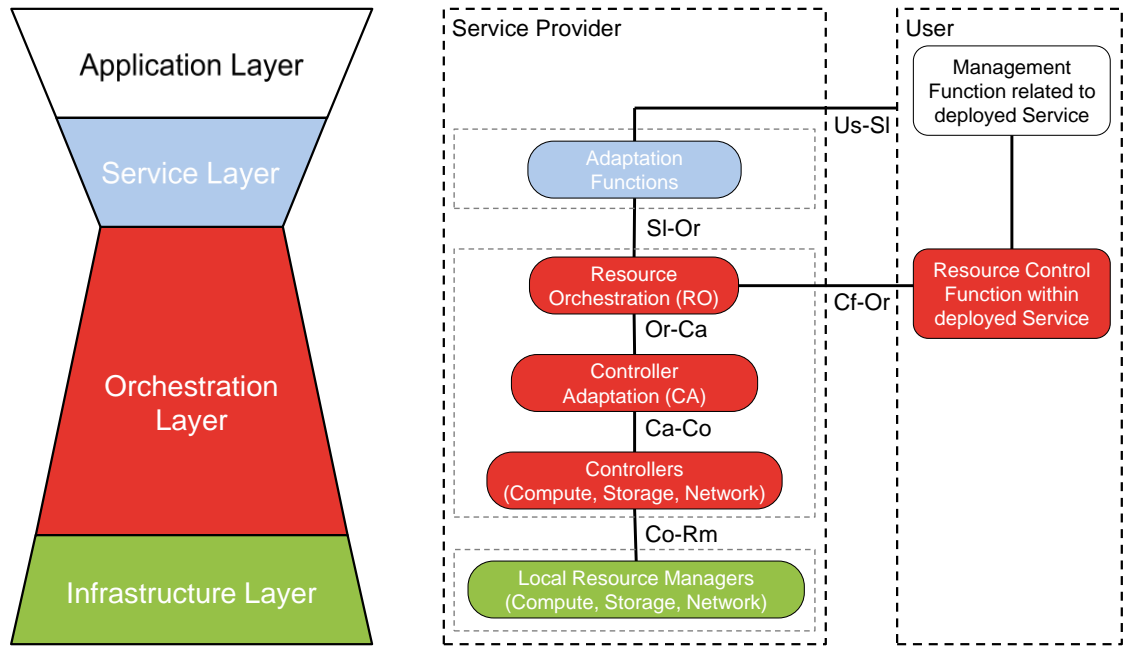

Figure 12: The three layered UNIFY architecture

The Infrastructure Layer encompasses all network, compute and storage physical resources. It can support the creation of virtual instances (networking, compute and storage) out of the physical resources. One of the challenges is to harmonize virtualization above these resources by proper abstraction in the Orchestration Layer. Four types of physical resources are identified:

i) our Universal Node (UN),

ii) SDN enabled network nodes (e.g., OpenFlow switches),

iii) Data Centres (e.g., OpenStack) and

iv) legacy network nodes or network appliances.

For development and prototyping, the main focus will be on the UN as it is the most novel type of resource developed within UNIFY. SDN enabled network nodes are also in scope given that they can fulfil the requirements for the specific observability and troubleshooting capability in question.

In the following, the 4 (sub)layers are depicted including the top-level functional model as described in detail in D2.1, representing the initial draft of the resulting functional architecture (Figure 13). The SP-DevOps concept and its related sub-processes (Observability, Verification, Troubleshooting, VNF-Development support) will mainly relate to the functional components on the right part of the sub-layers, which will be described in more detail in Section 4.2.2.

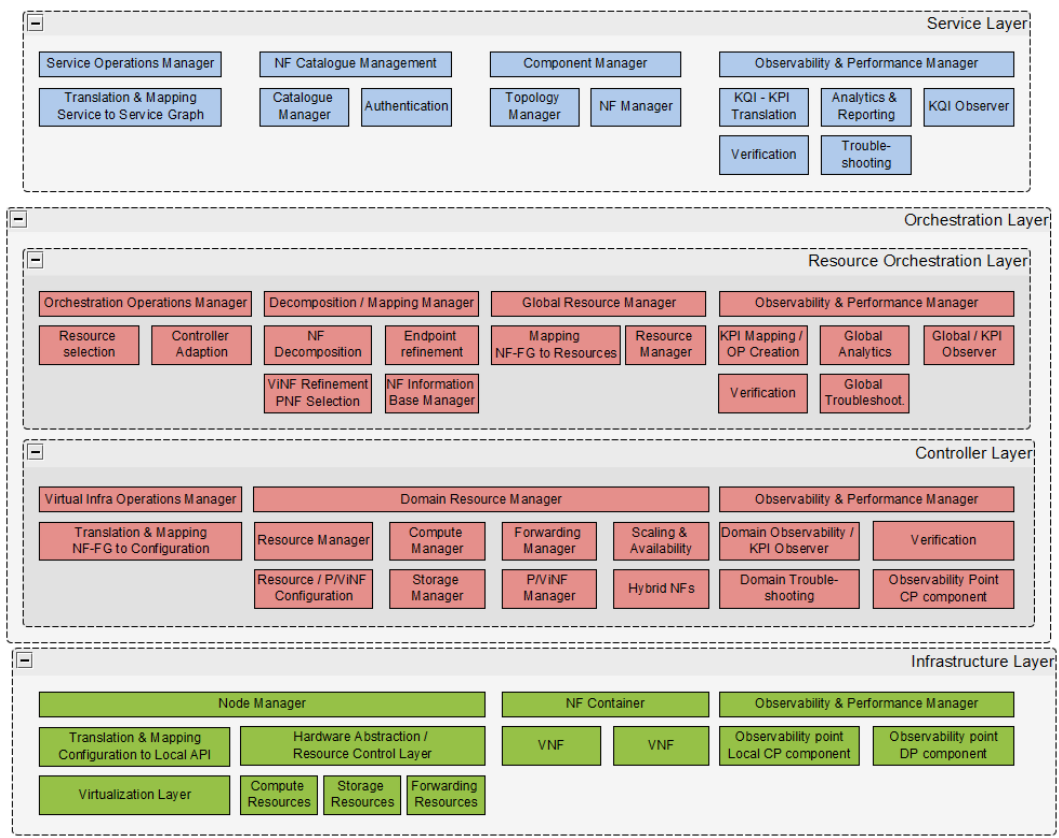

Figure 13: Initial UNIFY functional architecture

# 4 SP-DevOps concept

## 4.1 Sketch of SP-DevOps concept

In this section we define the roles of "developers" and "operators" in UNIFY. We will discuss briefly how we propose to apply IT DevOps principles in a carrier environment as envisaged in the UNIFY project. It has been widely acknowledged that a key component of DevOps is embedded in the cultural aspects that create an environment where development and operations team may interact with increased efficiency. We believe that this could be a topic for research in organizational processes and behaviours, but WP4 does not have the right capabilities in order to pursue such a research topic. Instead, we focus on the automation, measurement and verification aspects for DevOps, identify a set of related problems and present our first ideas on tools that could be developed to address these problems from a technical perspective. An organization may then adopt one or more of the solutions we propose and find the best way of integrating them in their own processes, along with defining how these processes will need to change to take advantage of the new technical capabilities. This approach is similar to the way typical DevOps tools such as Chef, Puppet and Ansible were developed from the ground up.

There are two facets of the "developer" role in UNIFY. One facet refers to the person that determines which high-level functions should be part of a particular service, decides what logical interconnections are needed between these blocks and defines a set of high-level constraints or goals related to parameters that define the service. This person might be the product owner for a particular family of services offered by a telecom provider. They might be a key account representative that adapts an existing service template to the requirements of a particular customer by adding or removing a small number of functional entities. We refer to this person as the *service developer* and for simplicity (access control, training on technical background, etc.) we consider the role to be internal to the telecom provider. The other facet of the UNIFY "developer" role is a person that writes the software code for a new virtual network function. Depending on the actual virtual network function being developed, this person might be internal or external to the telecom provider. We refer to them as *VNF developers*.

The role of the "operator" in UNIFY is to ensure that a set of performance indicators associated to a service graph are met when the service graph is deployed on virtual infrastructure within the domain of a telecom provider. In cloud computing companies such as Google, Amazon and Facebook, this role falls within the responsibility of the so-called site reliability engineering teams. From a standard telecom and enterprise management perspectives, this role is related to assurance processes in the eTOM framework and to the Service Operation in ITIL, respectively.

Compared to a standard cloud computing environment in which DevOps originated, a unified network and cloud set of resources in a telecom provider environment exhibits the following major differences:

- Higher spatial distribution, with lower levels of redundancy: telecom resources are spread over wide areas due to coverage requirements. In access and aggregation networks, the levels of equipment redundancy are much lower than in the massive data centres of typical cloud computing companies. Differences in latencies of any two control actions may trigger unsynchronized state changes between neighbour nodes, which in turn may translate onto legitimate packets being dropped in the parts of the network that have yet to receive the latest update to the flow table while incoming traffic is forwarded by the nodes that already received the updates.

- High availability is the norm, with stricter adherence to the four or five "9s" expected from customers. Infrastructure support for such capabilities needs to be verified at the service development stage, and downtime minimized during operations.

- Strictly controlled latency is required for many virtual network functions due to technical restrictions specified in standardization documents. This drives demand for more frequent measurements, impacts the precision required from measurement tools and requires automated handling of operations such as scaling or migration.

- A larger number of distributed data centres is needed [79] in order to address the spatial distribution and controlled latency requirements

The differences outlined above impact the way in which DevOps principles may be applied to telecom provider infrastructure. Although the principles remain the same, the potential costs of rolling back a service change are higher for telecom providers because redundant equipment might simply not be available to take over the functionality while a patch (be it software, or configuration) is being applied while high-availability requirements are expected to be strictly enforced. As such, we assert that an increased focus needs to be put on the verification and validation at various stages in the deployment and activation of service graphs by developers and operators. The ITU-T Y.1564 standard defines a methodology for validating standards-based Ethernet connectivity services at deployment time and continuously during their lifetime. We envisage that similar functionality and processes need to be developed for service graphs, although the problem in the service graph case is much more complex due to the huge increase in the potential number of parameters and verification tools. Having the right information on virtual network function and infrastructure metrics at the right place and at the right time has a high potential to reducing the troubleshooting time for problems. Finally, isolation capabilities for data plane and control planes in the network nodes and virtual network functions that enable testing in a production-like environment are paramount.

## 4.2 SP-DevOps applied on UNIFY

The UNIFY production environment plays a key role in the interaction between service or VNF developers and the operators. The UNIFY production environment provides the tools for orchestrating, configuring and controlling the virtual resources used by a service graph or virtual network function. This means that all the resources (in terms of compute capacity, network interfaces and tunnel identifiers for the dataplane, management plane and control plane connections and storage for both the infrastructure and the virtual network function management notifications) are identified and configured automatically as result of work performed in WP3.

The aim of SP-DevOps is to situate itself overall at the "Defined" level of the HP DevOps maturity model [11]:

- Customization of services or new releases of VNFs have the capability to be released frequently to the production environment based on policies provided by team members representing Operations. Our interpretation of release management from [11] is that it enables successive improvements of service graphs in a UNIFY production environment. Such improvements result by defining new categories or classes of services that include certain constraints or policies by default, or changing the composition of the service graph to include newly-developed virtual functions.

- Self-service one-click automated build, orchestration and deployment processes are identical in all environments (development, test, production). We expect to be able to go beyond "Defined" to the "Measured" level by providing visibility onto metrics related to the introduction of new services, such as release cycle time and defects.

- Collaboration is part of an established process and supported by a tool chain common to cross-functional feature delivery teams. In this area we will address only the common tool chain by providing tools that could be used by cross-functional team members addressing both development and operations scope. Cultural aspects, such as frequent communication characterized by mutual trust, cannot be addressed in this Work Package. Neither could we address detailed measurements of collaboration processes at the organizational level.

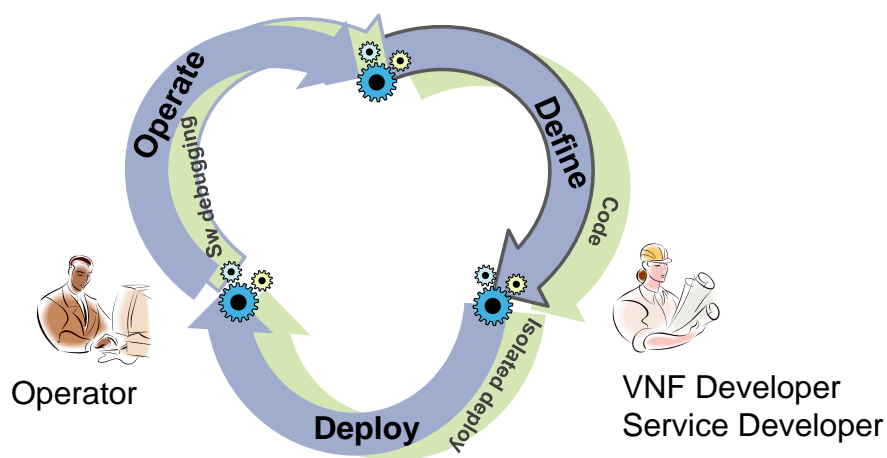

*Figure 14: Schematic representation of the SP-DevOps cycle*

In line with the principles of the IBM DevOps model [12], the observations regarding particularities of telecom provider environment outlined in Section 4.1 and the WP4 Description of Work, we define four large categories of processes that are integral part of SP-DevOps (i.e. Verification; Observability; Troubleshooting; Support for VNF Development). For completeness, we represent SP-DevOps in a cyclical form as depicted in Figure 14. The "Define" stage represents the process through which fulfilment functions determine which resources should be allocated to a service graph defined by a service developer. The "Code" stage is associated to the process of the VNF Developer writing the software that will be executed as a virtual network function. The "Deploy" stage represents fulfilment functions that configure and activate resources in the unified production environment. For the VNF Developer role, additional constraints, in particular in view of isolating the execution environment, need to be taken into account at this stage. This is therefore represented as an additional stage in the SP-DevOps cycle. Verification functions are the way WP4 implements this stage of SP-DevOps. The "Operate" stage is concerned with assurance functions related to the production environment and the VNF. For the VNF Developer role, this stage of SP-DevOps is concerned with debugging the software implementation in a realistic medium isolated within the unified production environment. Observability and troubleshooting functions are the way WP4 implements this stage of SP-DevOps. The VNF Developer may transition its virtual network function to the real production environment by making it available to the "Define" stage of SP-DevOps once the software debugging was successful – this is indicated through the blue outline of the "Sw debugging" arrow.

The cogwheels placed between stages of the SP-DevOps cycle symbolize the fact that automation is an inherent part of the concept. Automation is needed in order to respond to velocity and scalability requirements generated by the UNIFY production environment. Programmatic interfaces located at the transition between stages in the SP-DevOps cycle act as an enabler for the automation.

The four categories of processes dealt with in WP4 relate mainly to three out of the four DevOps principles [12]:

- <u>Monitor and validate operational quality</u>: This is the principle motivating the core activities in WP4: providing processes and accompanied methods and tools for *Verification* and *Observability*, together with capabilities supporting *Troubleshooting* of service graphs.

- <u>Develop and test against production-like systems</u>: With respect to development of new or updated virtual network functions (VNFs), we will detail a process supporting *VNF-Development*, allowing the developers of new or updated VNFs to deploy and verify their functions on the UNIFY production environment. With respect to development or definition of service graphs, the developers are supported by verification and debugging capabilities within the various layers of the architecture. And with respect to the operation of the infrastructure, isolation and virtualisation will be key concepts to protect other active services.

- <u>Deploy with repeatable, reliable processes</u>: This principle corresponds largely to one of the general UNIFY ideas about automatic deployment and operation of service graphs. While the fulfilment related processes are part of WP3 (Service Deployment process and programmability framework), reliability of these processes is supported by programmable and automated observability and verification capabilities developed in WP4.

The forth principle <u>"amplifying feedback loops"</u> is very much connected to both the culture within an organization, as well as business decisions based on market or customer feedback, and is as such not directly covered by the technical solutions developed in WP4. However, this principle is obeyed in UNIFY from a technical point of view by

35                                                          Deliverable D4.1                                                          10.02.2015

providing a customer facing service layer, including interfaces towards fulfilment and assurance processes, potentially even development processes.

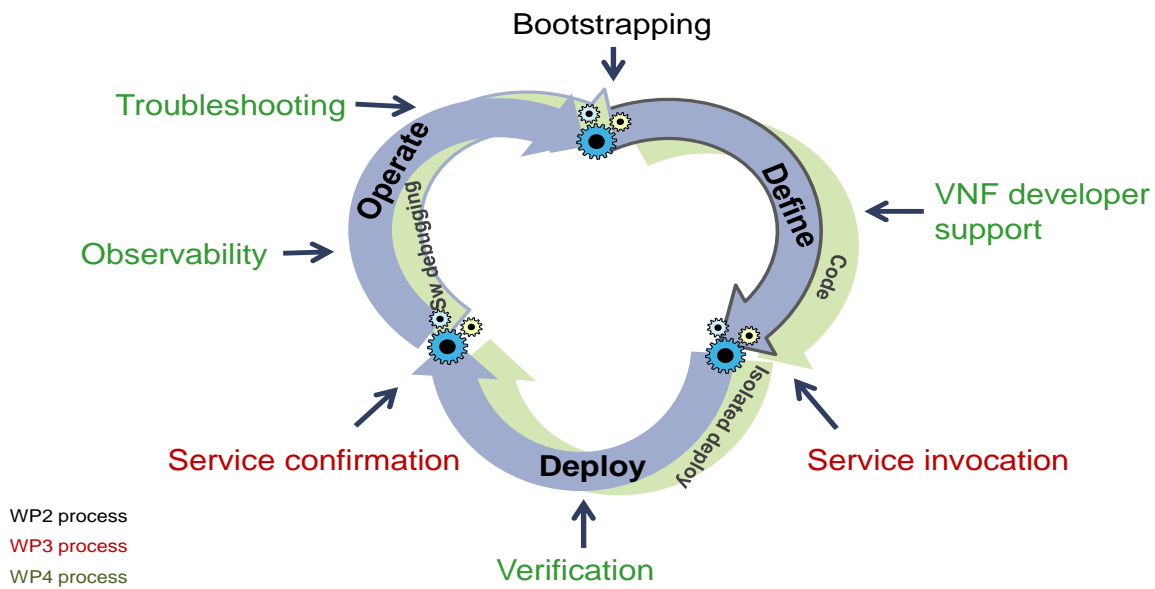

Figure 15: Relations between SP-DevOps stages and UNIFY processes

Figure 15 depicts the relations between processes currently defined in D2.1 (the WP4 processes are further detailed in Section 4.2.2) and the overall SP-DevOps stages. The bootstrapping process (D2.1) is the base for adding new virtual functions and resources. The Service Invocation processes (D2.1) enables the Developer (whether for a service, or a network function) to define a service graph and automatically trigger its deployment onto suitable resources. When the deployment is completed, the Service Confirmation process (D2.1) announces the resource identifiers and enables operations capabilities to either assure the day to day operations or participate in a distributed virtual network function debugging session.

Before describing the four large categories of SP-DevOps processes in relation to the functional architecture, it is important to understand our definition of monitoring functions and their observability components envisioned for UNIFY. This is relevant, since most of the processes and research questions developed in the scope of UNIFY WP4 will be centred around observability and troubleshooting of service chains and its building blocks.

### 4.2.1 Definition of Monitoring Functions

Since monitoring and validation of operational quality has been identified as a key principle of SP-DevOps, we want to give our definition of monitoring functions and corresponding components. The following definitions are embedded into the UNIFY functional architecture (Figure 13) and are described separately due to their importance for the remainder of this deliverable.

The overall purpose of the monitoring functionality implemented in UNIFY is to increase the observability over the network behaviour and state, as a trade-off between scalability (i.e. resource-efficiency in terms of monitoring overhead) and accuracy. Certain monitoring functionality will also address the unification of network and compute resources.

Monitoring functionality is implemented as Virtual Network Functions (VNFs) operating in both control and data planes. Conceptually, such a VNF consists of one or several observability points (OP) instantiated on one or several virtual nodes. The monitoring function consists of a control plane (CP) component for analysis and control of lower-level monitoring operations towards the observability points. In turn, an observability point operates in terms of a local CP and data plane (DP) components for local analytics and measurement purposes (Figure 16).

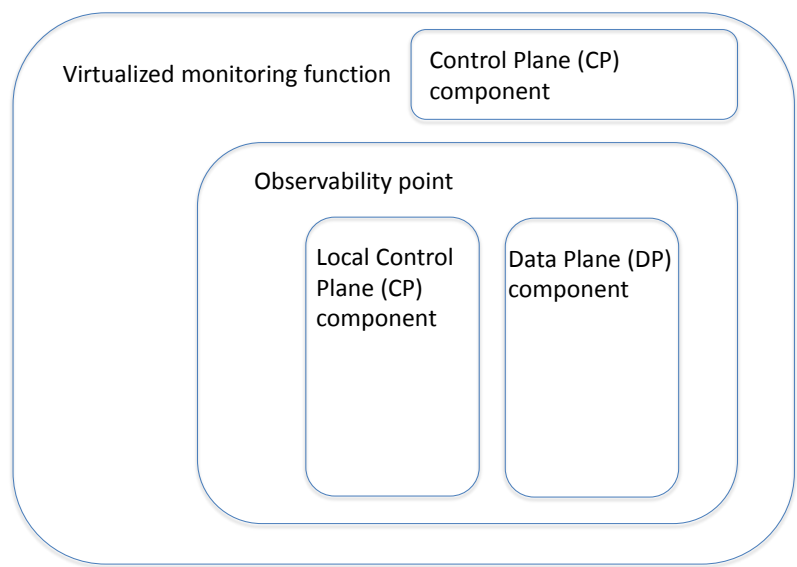

*Figure 16: Conceptual overview of a monitoring function.*

The operational scope of a monitoring function includes the Virtualized Infrastructure Management Layer (Controller Layer) and the infrastructure layer. A monitoring function implements different levels of observability and analysis for performance monitoring and troubleshooting purposes. The complexity of the monitoring function can vary from very simple probing mechanisms in the infrastructure layer to more complex analytic applications based on one or several observability points, instantiated in one or several nodes.

The monitoring function includes a CP component that implements mainly three functions:

- control domain and global analytics of reported measurements;
- notifications for further performance analysis or troubleshooting purposes;

- dynamic control functionality towards certain OPs in the infrastructure layer.

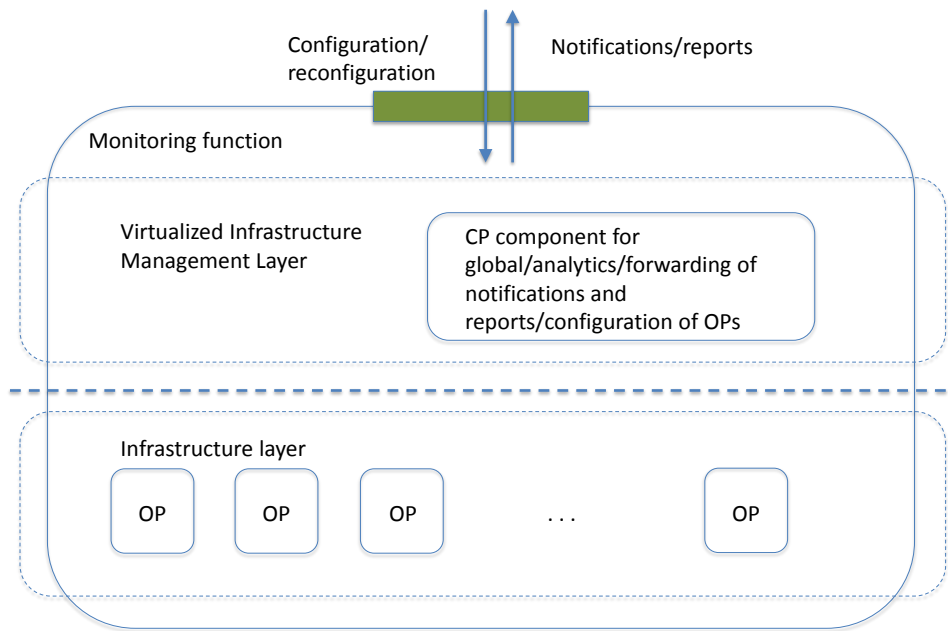

Figure 17: The main operational scope of a monitoring function encompasses implemented functionality mainly in the Controller Layer and infrastructure layer.

The definition of a monitoring function enables the implementation of monitoring capabilities at varying levels and complexities. A simple monitoring function can, for example, be implemented for monitoring delay and loss on a certain link (physical or logical) relative to specified QoS requirements. The delay and loss metrics can be observed by instantiated observability points in two or more nodes involved, and modelled using e.g. probabilistic approaches [80]. Depending on the implementation, the link metrics can either be provided in terms of separate observability points for delay and loss, or from the same observability point. A more complex monitoring function can for example implement trend analysis based on model estimates collected from several observability points in the network.

The monitoring function is instantiated via an interface of the virtualized function by the orchestrator given configuration and monitoring specifications (e.g. type of monitoring function, requirements on accuracy, detection thresholds, etc.). The instantiation in the Controller Layer includes mapping of the observability points on to certain nodes in the infrastructure layer. The number of observability points and the placement of them are parameters specific to the implementation of the monitoring function, but the exact mapping onto resources in the infrastructure layer is part of the resource management components in the orchestration layer.

### 4.2.1.1 Observability points and components

Observability points implement measurement and modelling functionality local to a virtual node and consist of two types of components operating in the local CP and the data plane (DP) (Figure 18). The output from one or several observability points can be used for high-level analytics in the monitoring function at the Controller Layer.

In general terms, typical for the local CP component is low frequency updates and control operations and high complexity computations (updates of estimates, local change detection, troubleshooting support, etc), whereas the DP component operates at a high frequency and low complexity (counter updates, flow entry matching, etc).

Local CP components enable in-network analytics, which is an important part in meeting the requirements on scalability, specifically for measurement intensive monitoring functions. The main tasks of the local CP component are to:

- execute control operations relevant to specific implementations of a monitoring function, such as scheduling of active measurements, packet manipulation in certain cases when an active measurement tool is implemented with CP support, message and information exchange between OPs, or forwarding of certain packets to the Controller Layer CP component;

- perform high-level computations for the purpose of node-local performance analytics of e.g. a certain flow or flow aggregates, that can be further used for e.g. detecting local performance degradations and other types of troubleshooting support;

- act as a measurement data transport intermediary, that pre-aggregates or pre-processes data in the network to limit the bandwidth and computational load on the analytics component of the virtual monitoring function;

- report the observed behaviour based on the fulfilment of specified conditions, or in terms of regularly pushed messages in accordance with specification, or by request. The receiver of such a report or notification can be a developer, service user, operator or another management function (e.g. for dynamic resource management purposes).

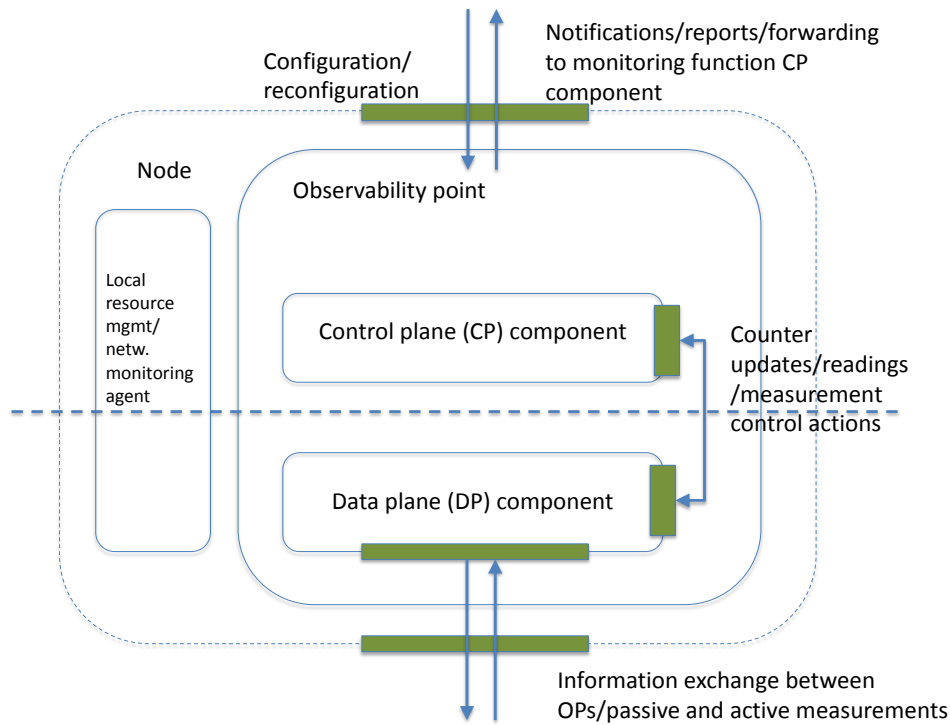

*Figure 18: Overview of an observability point and components in a virtual node. The observability point can be configured and reconfigured by a specific interface, but the instantiation of an OP is handled by a local resource management processes in the node.*

Note that the degree to which a local CP component is implemented in each OP is dependent on whether the monitoring function is implemented for centralized or in-network monitoring. For example, a centralized monitoring function strictly following the SDN paradigm would for the largest part be based on the centralized CP component in the Controller Layer for performance monitoring and analytics, rather than on a local CP component. Thus, a local CP component can be very simple in terms of forwarding a packet from the DP component to a centralized CP component, or perform more complex operations such as producing estimates of the observed network behaviour, or send a test packet at certain intervals, for example.

The DP component performs low-complexity operations but at a high frequency specific to the implemented monitoring functionality. Such operations mainly include:

- passive measurement support, such as counter updates to specified flows or flow-aggregates;
- active measurement support, such as packet manipulation (e.g. timestamping and marking) and creation of measurement probes;

- packet forwarding to the local CP component.

Operations performed in the DP component are supported by the capabilities of the DP provided by the UN, which are further outlined in the WP5 deliverable D5.2 [81] It is expected that the DP of the UN supports OF-counters but also allows for software-defined counters that can perform simple arithmetic operations, in order to observe different aspects of the network behaviour (e.g. at different time granularities). This is necessary for the support of dynamic and flexible network operations. The degree of observability, resource-efficiency and accuracy in modelled link metrics and monitored network behaviour depends on the frequency with which the operations in the DP component are performed (e.g. the frequency of counter readings), as well as the resources needed to perform certain actions (e.g. number of counters needed to accurately model a flow behaviour).

#### 4.2.1.2 Interfaces

Upon instantiation of a service graph, the monitoring functions are specified in accordance with desired accuracy and the conditions under which the monitoring functions should operate. This specification can include deterministic measurement rates, detection thresholds, or input specifying probabilistic guarantees on the monitoring performance. From a resource-consumption perspective, it is assumed that the orchestration layer performs functional decomposition determining the monitoring objectives relevant to each layer of the composition, as well as resource management functionality for instantiating a monitoring function and associated observability points.

A monitoring function consists of a set of interfaces that allow for instantiation and (re-) configuration of the monitoring behaviour as well as for notification to the Controller Layer and the orchestrator (Figure 17). A Virtualized Network Function interface allows for configuration and instantiation of each observability point from the Controller Layer CP component specific to each monitoring function (Figure 18). Moreover, OPs instantiated at the nodes can exchange messages (such as test packets) for measurement and troubleshooting purposes (Figure 18).

### 4.2.2 SP-DevOps process flows

We will in the following describe the four large categories of processes in focus of UNIFY WP4 (Observability, Verification, Troubleshooting, VNF Developer support,) in more detail and map their functionality to each UNIFY architecture layer and involved components described in D2.1. Figure 19 illustrates a simplified overview of the functional architecture with the components within the main scope of WP4 highlighted.

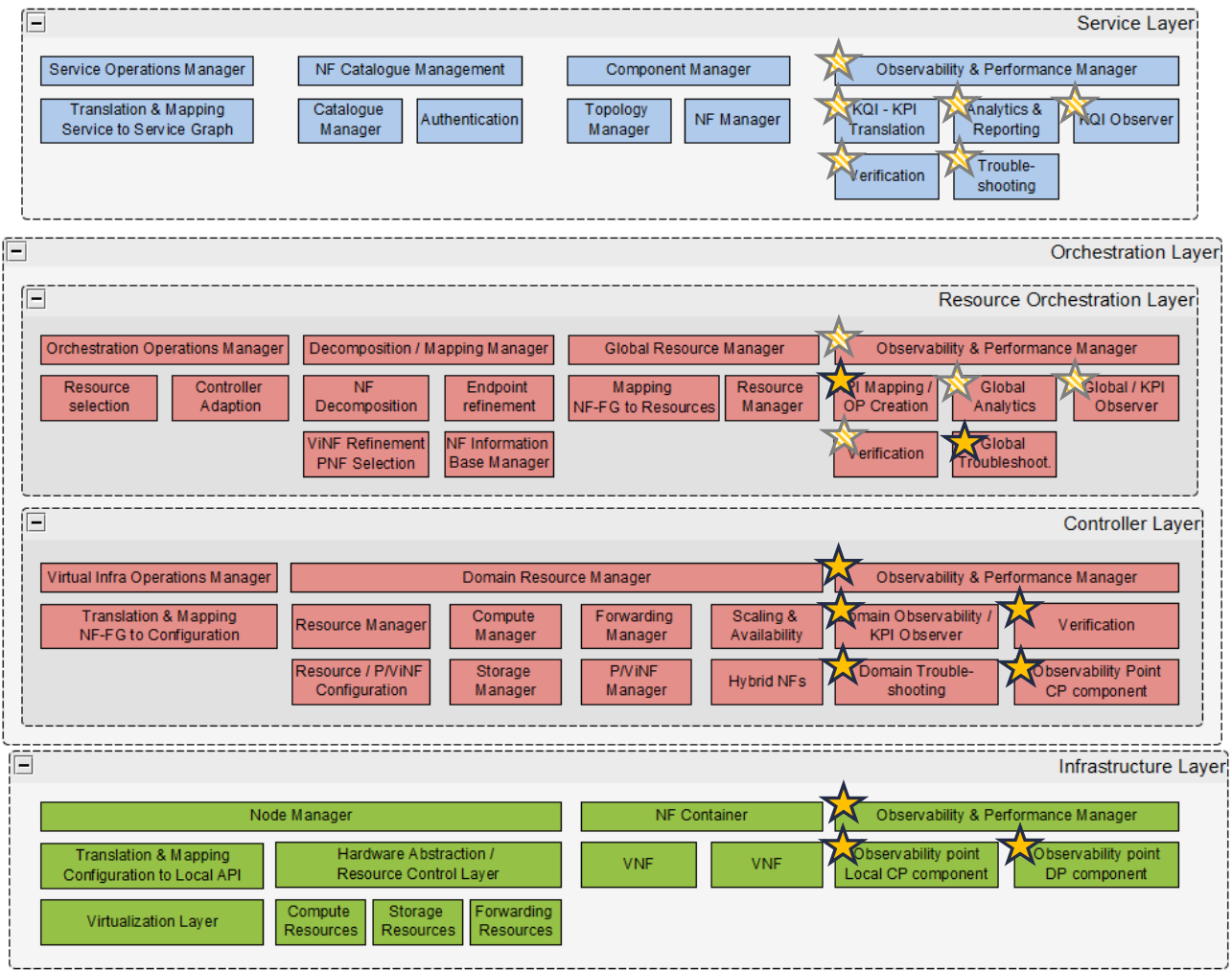

*Figure 19: Functional UNIFY architecture with WP4 focus: Solid stars represent identified UNIFY DevOps research challenges (Section 4.3); stripy stars represent integration aspects.*

Although we present the high-level processes views here, it is understood that the major research contribution will be in terms of technical aspects covered by particular parts of these processes. The processes are important from the perspective of integrating the output of the Work Package with WP3 and WP5 in the integrated testbed. Identifying them in the beginning of the project gives us a communication tool towards the other Work Packages and contributes to understanding what interfaces and common functionality would need to be developed in order for the integration to take place.

#### 4.2.2.1 Observability process and associated functional components

The Observability process provides visibility onto the operational performance of service graphs deployed in the unified production environment. The process (Figure 20) has two distinct objectives:

1. Key Performance and Quality Indicator (KPI/KQI) translation and tool selection which determines what should be measured for a service graph, how such measurements are carried out and where such measurement capabilities are instantiated and configured. This is represented in the figure (depicted as yellow arrows) as the chain of events going from a Developer down through the layers terminating at the Infrastructure layer.

2. Measurement data generation and analysis which rely on novel capabilities for performance and fault management developed by UNIFY partners. Basic data generation capabilities on any type of resource include e.g. packet/byte counters, notifications and logs generated by the resource itself or a function executing on the resource. This is represented in the figure (depicted as green arrows) as the chain of events going from the Infrastructure layer up to the Developers.

Starting with the first objective, KPI/KQI translation and tool selection, there are several functional blocks in the architectural layers to fulfil this objective:

- In the **service layer** an incoming service graph contains an SLA that describes a set of KPIs and/or KQIs that need to be fulfilled during the lifetime of the service. These KPIs and KQIs provide the first trigger of the SP-DevOps Observability process by dictating which measurements are needed for a service graph and how they are analysed. In the service layer these need to be translated and mapped from high-level abstract KQIs that may be applied to compound network functions to more precise KPIs by the "KPI – KQI Translation" functional component.

- In the **orchestration layer** the KPIs that are provided from the service layer as part of an NF-FG are mapped to Observability Points, which in turn are mapped onto virtualized infrastructure management and infrastructure layer resources that can provide the required measurements and analysis to fulfil the KPIs. This is done by the "KPI mapping / OP creation" functional block, whose output is taken into account during the orchestration process. In the resulting instantiable NF-FG produced by the orchestration layer new NFs may be included to provide the necessary observability components, if their corresponding KPI calculation cannot already be provided by the existing infrastructure resources or existing NFs. Additionally, the orchestration layer includes a second trigger in the SP-DevOps Observability process when it requests the creation of Observability Points that are not the direct result of service graph specifications, in order to obtain up-to-date resource utilization and infrastructure health monitoring data. The orchestration layer uses this data as part of the orchestration process in order to take the current state of the network and compute resources into account when mapping NF-FGs to the available resources. These triggers are created by the "Resource manager" functional block.

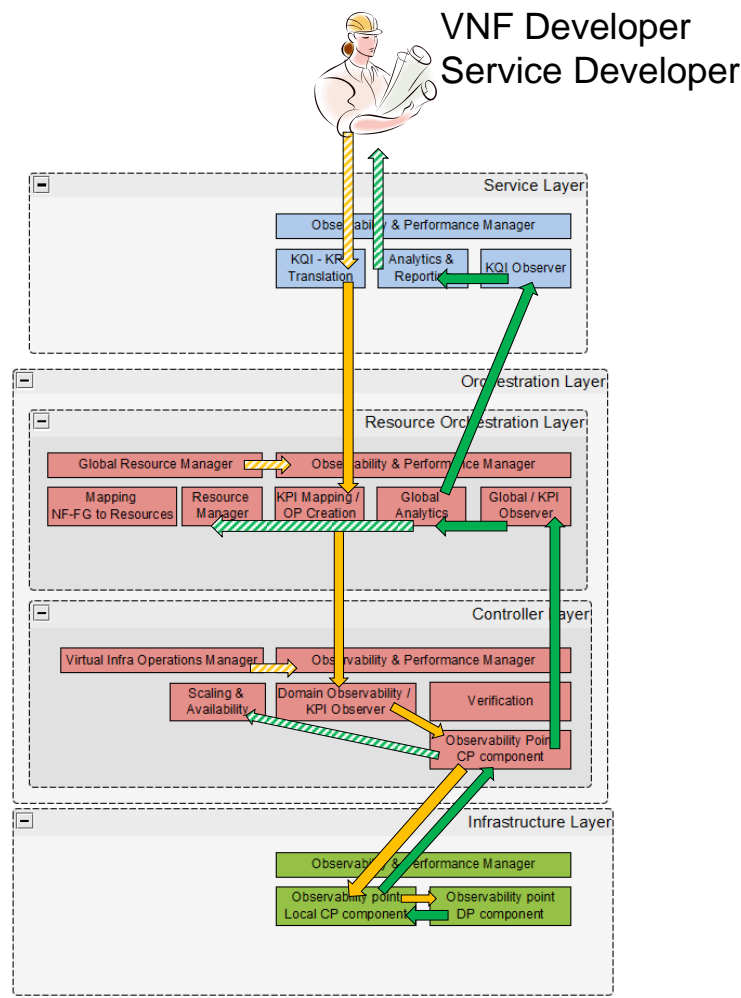

*Figure 20: The Observability process in relation to architectural layers: Yellow arrows indicate the observability request and invocation; green arrows indicate notification and reporting; solid arrows represent the parts of the process dealt with in WP4; dashed arrows represent the parts of the process flow that need to be done in shared responsibility with other WPs.*

- At the **Controller Layer** the KPIs that are part of the incoming NF-FG are mapped to existing Observability Points in the infrastructure layer or new one are allocated. This is done by the "Domain observability / KPI Observer" functional block together with the Domain resource manager, in particular the "Resource Manager" function within the Domain resource manager. Observability Points in this layer are represented by an "Observability Point CP Component" block which is responsible for coordinating the lower layer Observability Points instances and for aggregating/analysing their output. The output of Controller Layer is then communicated to the infrastructure layer using technology specific protocols for the particular infrastructure.

- In the **Infrastructure layer** Observability Points for gathering measurements and performing initial analysis and aggregation are instantiated and configured based on commands from the virtualized infrastructure management layer. Observability Points are represented by two functional blocks, "OP local CP component" and "OP DP component", the first performing aggregation/analysis at the node level and the second representing the functionality performing the actual measurements. The functional block responsible for instantiation and configuration of these functions is the "Observability & Performance Manager".

For the second objective, measurement data generation and analysis, there are also several functional blocks in the architectural layers that are involved, starting from the infrastructure layer:

- In the **Infrastructure layer,** data is gathered, analysed and aggregated by the "OP local CP component". Analytics results are then forwarded to the Controller Layer for further analysis. Analytics performed in the "Observability Point CP component" includes e.g. modelling of flow counters and modelling of measurements such as delay or loss that might be performed in the node. The "OP local CP" component also generates notifications/alarms in case of e.g. failures, which are sent to the Controller Layer where management decisions may be taken. Such alarms may also have a role in the Troubleshooting process, detailed in Section 4.2.2.3.

- The **Controller Layer** obtains locally aggregated/analysed measurements and other data provided by observability points in the infrastructure layer and performs domain level analysis, using the "Observability Point CP Component" functional block. The received data is also used in the "Domain observability / KPI observer" functional block in order to monitor the KPI status versus thresholds and generate notifications when thresholds are breached. Results of these functional blocks are sent to the orchestration layer, and/or acted upon within appropriate layer via local triggers.

- In the **Orchestration layer** notifications and modelled data is received for further analysis and troubleshooting/resource management support, the data is processed in the "Global analytics", "Global observability / KPI Observer" functional blocks. Relevant data is provided to the "Resource manager" functional block for updating the global resource view and informing the orchestration process. The "Global analytics" and "Global observability / KPI Observer" also sends notifications and KPI status updates to the Service layer.

- Finally, in the **Service Layer,** notifications and KPI statuses are integrated and translated back to the original KQI/KPIs requested in the service graph(s) by the "KQI-KPI Translation", "KQI Observer", and "Analytics & reporting" functional blocks.

Moreover, several of the steps in the Observability process are required to be dynamic in the sense that they are expected to be triggered following the increase or decrease of resources allocated to a service graph, when service graph components are migrated through the production environment.

### 4.2.2.2 Verification process and associated functional blocks

Enabling ongoing verification of code is an important goal of continuous integration as part of the DevOps concept. While traditional DevOps mainly refers to verification of code, we relate this goal in SP-DevOps to verification of service definitions and configurations. Automated verification functions on each layer of the architecture facilitate verification as part of each step in the deployment process, allowing identification of problems already early in the service lifecycle. In that sense, verification is less of a process, but rather a set of features providing gatekeeper functions to verify the abstract service models - Service Graph, Network Function –Forwarding Graph (NF-FG), and the proposed resource configuration - before actual instantiation on the infrastructure layer takes place. The Verification process is outlined graphically in Figure 21.

We describe the Verification process and its functionality mapped to UNIFY functional architecture from Figure 19):

- The **Service Layer** receives a service request from developers in the form of a Service Graph with a related SLA definition. This abstract Service Graph definition will allow the SL to verify the absence of loops and other topological consistency properties. As a result, Service Graphs can be marked as invalid and returned to the customer/user early on in the deployment process. If the verification does not find any inconsistencies, the deployment process continues in the next layer. The result of the Service Layer is a NF-FG which is also verified for absence of loops and other topological properties before it is sent to the orchestration layer.

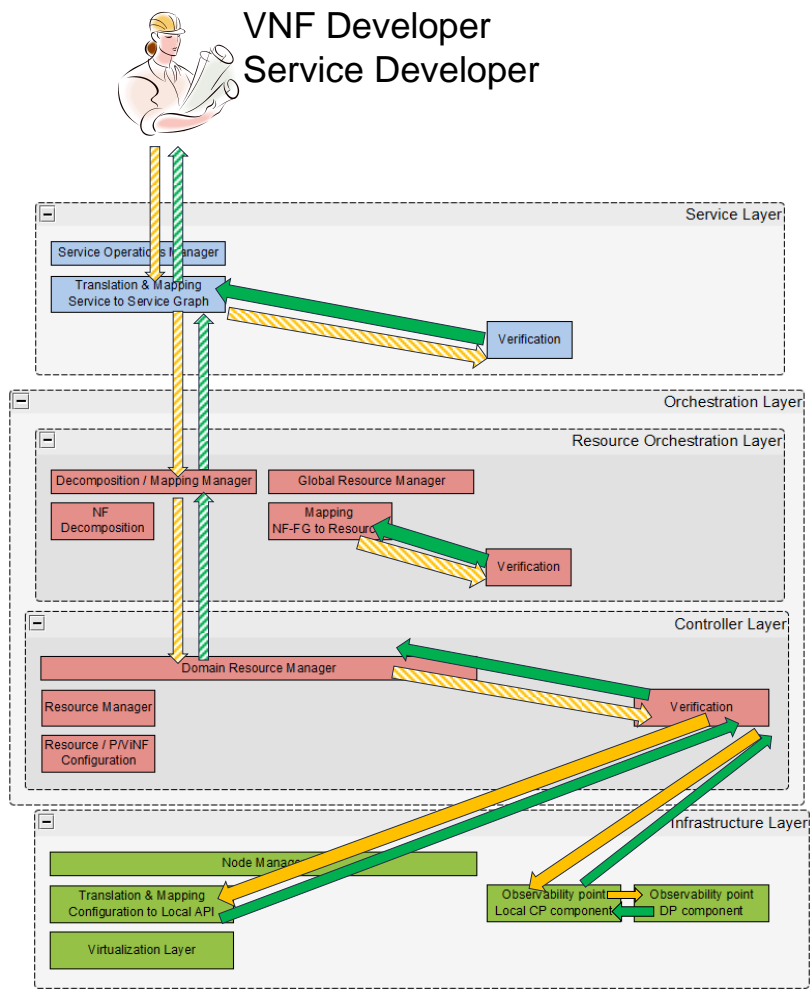

*Figure 21: Verification process in relation to architecture layers: Yellow arrows indicate the validation requests and invocation; green arrows indicate notification and reporting; solid arrows represent the parts of the process dealt with in WP4; dashed arrows represent the parts of the process flow that need to be done in shared responsibility with other WPs.*

- The **Orchestration Layer** receives a NF-FG and performs the placement of the contained NFs (i.e. finds locations where they can be instantiated or already running NFs that can fulfil their requirements), resulting in an instantiable NF-FG which can be verified against consistency with respect to the resource, capability and topology descriptions. Furthermore, on this layer the instantiable NF-FG can be verified against policy violations related to placement of NFs and performance impact on already deployed NF-FGs on the chosen infrastructure.

- The **Controller Layer** consists of a set of different controllers for both compute and networking resources. The verification functionality in this layer targets consistency of specific configuration instances, such as inconsistent network configuration in the form of OpenFlow rules. Verification of the services and service components instantiated in the **infrastructure layer** are functionally handled in the virtualized infrastructure management layer, following SDN principles. Therefore we have not planned any verification or validation functionality directly in the infrastructure layer. The verification functionality for the infrastructure and the virtualized infrastructure management layer is implemented in the "Verification" functional block.

#### 4.2.2.3 Troubleshooting process and associated functional blocks

With troubleshooting, we mean the localization of the source of a problem (the "trouble") related to a certain process. Troubleshooting mechanisms will need to operate on several levels of the architecture and will largely take advantage of the Verification and Observability mechanisms and tools introduced as part of the processes described above. By methodically ruling out potential causes, the actual root cause can be identified.

A troubleshooting process is triggered either manually by a developer, or by automated Service and Orchestration layer components based on reports or notifications provided by deployed Observability points. A requested troubleshooting process aims to follow up on reported bugs, faults, and anomalous states that require further investigation when the faulty or anomalous condition cannot be immediately localized from existing observations and reports. This includes automated deployment and re-deployment of relevant verification and Observability points in order to isolate the root-cause of detected bugs, faults and anomalies.

Fault localization reports are also generated asynchronously in the infrastructure layer, as in-network troubleshooting capabilities in the infrastructure layer run autonomous fault localization and root-cause analysis by analysing exchanged information (estimates, log or audit information, states, counter values, etc.). Detected and localized faults and performance degradations in the infrastructure layer are reported to the virtualized infrastructure management layer (and forwarded if necessary to higher layers) where further investigation may happen, or other tools may be triggered etc. This differs from manual triggers or those generated by higher layers as the fault localization may already be completed by the infrastructure itself.

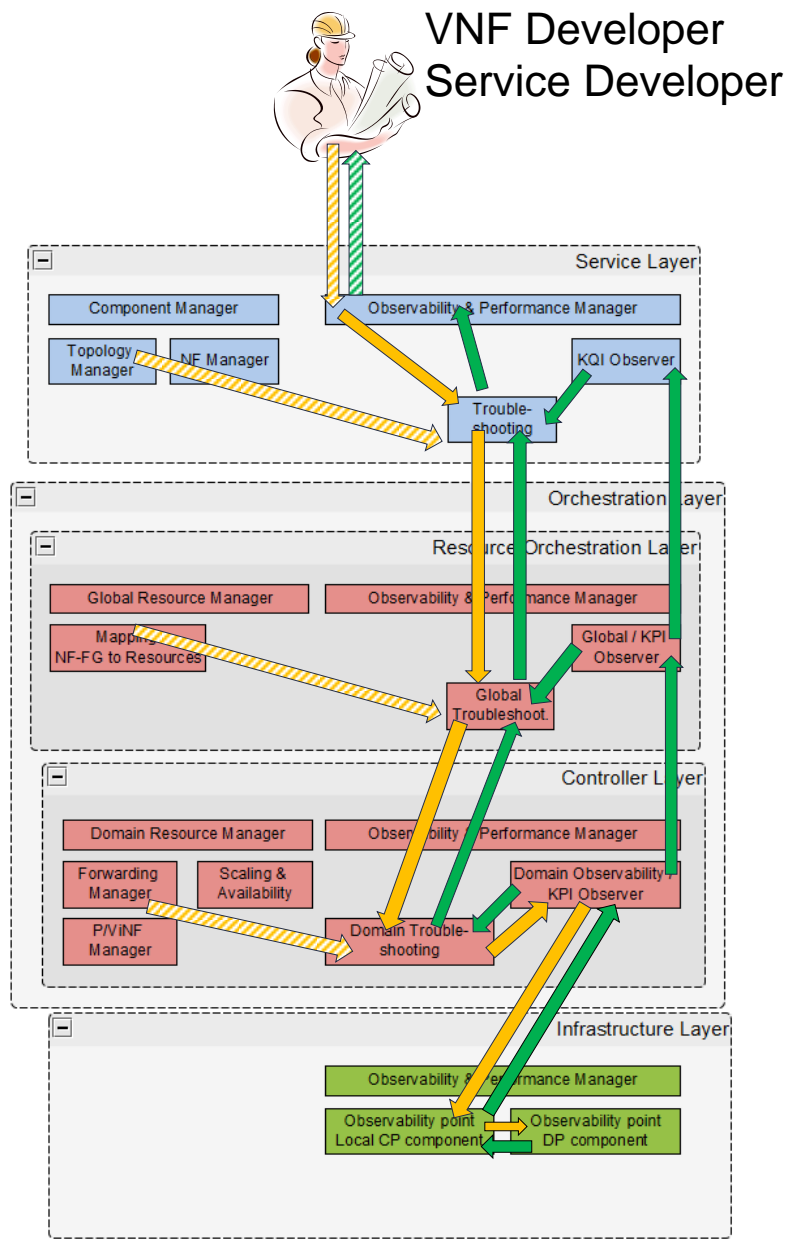

*Figure 22: Troubleshooting process in relation to architecture layers: Yellow arrows indicate the troubleshooting requests and invocation; green arrows indicate notification and reporting; solid arrows represent the parts of the process dealt with in WP4; dashed arrows represent the parts of the process flow that need to be done in shared responsibility with other WPs.*

The troubleshooting process depicted in Figure 22 shows the different ways troubleshooting / fault detection may be triggered:

- The **Service Layer** participates in the troubleshooting process mainly as a conduit of troubleshooting requests and by providing an interface to the troubleshooting capabilities for developers. This is done by the "Troubleshooting" functional block in the service layer.

- The "Global Troubleshooting" functional block in the **Orchestration Layer** receives triggers from layers above and below to the "Global Troubleshooting" functional block, as well as triggers from other functional blocks in the same layer, the "Global Analytics" and "Global / KPI Observer" blocks. The "Global Troubleshooting" block in the orchestration layer is also responsible for automating various troubleshooting tasks using measurements in the orchestration layer and troubleshooting capabilities of lower layers (e.g. by triggering the creation of new Observability points and analysing their results). The "Global Troubleshooting" block is also responsible for presenting these automated workflows as troubleshooting capabilities to the Service layer above it.

- The "Domain Troubleshooting" block in the **Controller Layer** serves a very similar role as its counterpart in the Orchestration layer but on a domain level, typically without a global view of the system. It receives triggers from layers above and below, as well as internal triggers from the "Domain Observability / KPI Observer" and "Observability point CP component" functional blocks. It also makes available domain-level troubleshooting capabilities to higher layers for automating troubleshooting workflows.

- The **Infrastructure Layer** provides troubleshooting capabilities to the higher layers mainly in the form of Observability points (the "Observability point CP component" and "Observability point DP component" functional blocks) and Verification tools (in the "Verification" block) that can be created and configured by higher layers. Some of these Observability points may additionally perform in-network fault localization and thus independently locate faults and report these to higher layers.

### 4.2.2.4 VNF Development support

When applied in the unified production environment, the VNF Development process supports the team developing functionality of a network function in line with the DevOps principle "Develop and test against production-like systems" mentioned in Section 4.2. The unified production environment provides the means to instantiate a newly developed or updated VNF onto the infrastructure. It also provides the means to identify the resources where a particular instance is being executed, which is important for debugging purposes. Note that the VNF Development process includes only the interactions with entities belonging to the unified production environment, i.e. neither considering the Developer's IDE nor functionality of traditional OSS/BSS systems (a discussion about the split between orchestration and management can be found in D2.1). For example, the actual copying of the code to the production environment, as well as configuration tasks to be performed before and after the instantiation, are part of the interaction with the OSS/BSS and thus not depicted in our process.

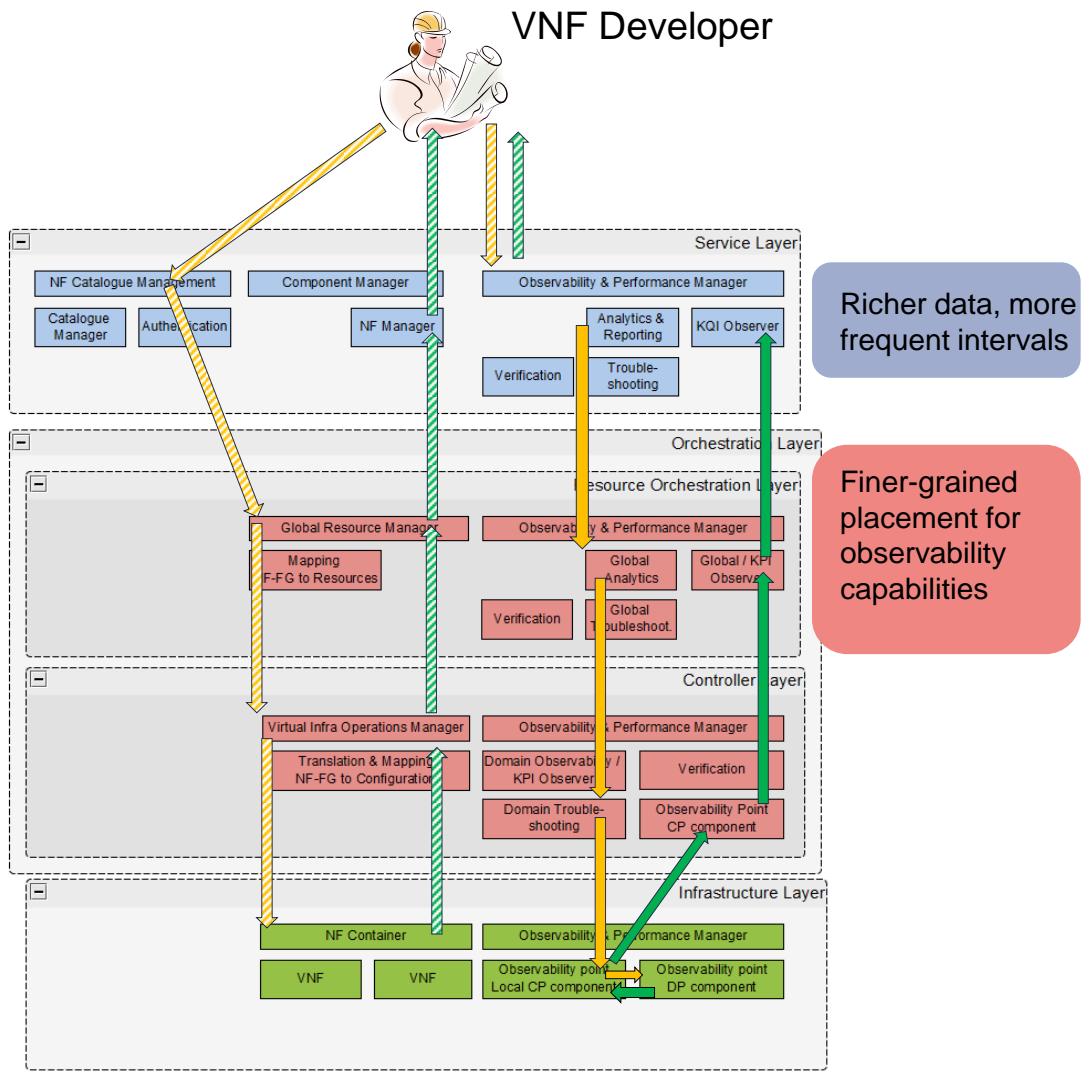

Figure 23: Processes supporting Development of VNFs: Yellow arrows indicate the downstream requests; green arrows indicate upstream responses; solid arrows represent the parts of the process dealt with in WP4; dashed arrows represent the parts of the process flow that need to be done in shared responsibility with other WPs.

In order to facilitate VNF development to be performed directly in the production system, there needs to be a set of supporting functions by the architecture towards the VNF developer on top of the Service Layer. We consider three sub-processes which will provide this functionality. Once a VNF under development is deployed within the production system, VNF developers will be supported with the observability, verification and troubleshooting capabilities described in the preceding subsections. During the VNF development process, we envision the observability capabilities for the specific network function to be placed more extensively in the infrastructure (finer-grained placement) and produce data in higher frequencies (as depicted as the right-hand flow in Figure 23).

The three additional sub-processes are described below, and are depicted as the left-hand flow in Figure 23:

- Adding a new VNF to the production environment: This sub-process allows developers to add their new VNF (or a new version of an existing VNF) to the production environment for testing and debugging purposes. To start with, the developer submits a description of the VNF capabilities and resource requirements to the Service Layer, which in turn informs the Orchestration about the existence of the new VNF. The Service Layer also stores the description of the new VNF in the service catalogue or inventory. The "adding a new VNF" view is also employed when updating an existing VNF. In this case, the updated VNF is stored in the service catalogue and the Orchestration is informed about any changes in terms of resource requirements that might have been introduced by the update. Optionally, if the resource model of the VNF is not yet known, then the Service Layer forwards a request to the Orchestration layer which determines by creating a simple service graph what are the resources allocated to the sample instance of the VNF.

- Modifying an already deployed service graph with a new or updated VNF: Besides debugging isolated VNFs, developers can proactively minimize bugs by testing their new VNF code embedded in a complete service graph, interacting with other VNFs used in the production environment. Here, the developer announces their intention to deploy a particular VNF (identified through a VNF id) in a particular service graph (identified through a Service Graph id) instantiated in the unified production environment. The request is received at the Service Layer, forwarded to the Orchestration layer which queries the service catalogue for the resources allocated to the current instance of the VNF and what resources need to be allocated to the new instance. The Controller Layer is responsible for allocating the new resources and it configures the policies associated to steering traffic towards the new VNF instance. Once the policies are in place, the developer is informed about the availability, identifiers and possibly location of the resources and can proceed with the actual copying of the binary VNF distribution, which takes place outside the VNF Development process.

- Attach VNF to software IDE: This sub-process is an enabler for actual VNF debugging activities. The developer queries the Service Layer by providing a Service Graph identifier and the type of the VNF in order to determine where the instance they are interested in is being executed. The Service Layer forwards the request to the Orchestration, which determines what resources are used for the VNF instance and provides the developer with an identifier for the resources and the VNF instance. The developer can then connect to the running VNF instance by means of tools such as distributed software debuggers, developed outside of UNIFY, and perform the debugging activity. The assumption of the "Attach VNF to software IDE" view is that no special needs have to be fulfilled in order to enable to connect to the VNF using a software debugger. In cases when special functionality needs to be deactivated or modified (for example, when the VNF instance is protected by a firewall), the developer needs to submit to the Service Layer a request to change the service graph accordingly.

## 4.3 Research challenges and proposed tools

After the generic description of the SP-DevOps related processes targeted by WP4, the following subsections will highlight specific research questions identified within the scope of the various processes. We will describe the challenges in connection with first indications of the tools and methods that are planned to be investigated and developed in WP4 Task 4.2 and 4.3, and eventually implemented in the SP DevOps prototyping Task 4.4. In Annex 2 we present a mapping of the research challenges against the objectives defined for the Work Package in the Description of Work document.

### 4.3.1 Observability

The UNIFY service provider will depend on the ability to monitor fault- and performance metrics for various reasons, ranging from SLA assurance to real-time and in-depth observability needs supporting increased velocity and dynamicity of tasks such as network planning, performance analysis and optimization. However, programmable monitoring capabilities will not only support the operational aspect of a provider, but also support any type of *service developer* (which might be a role within the providers organization, a role within a trusted partner of the provider, or in the future even a 3$^{rd}$ party developer) with greater visibility into the performance characteristics of the virtualized infrastructure, thereby facilitating verification, troubleshooting and service performance evaluation.

A major problem related to frequent and fine-grained observability updates from many components, as envisioned in UNIFY, is scalability and resource-efficiency. In WP4, we will study and propose multiple complementing approaches to tackle this challenge, including distributed- and centralized- (i.e. controller-based) solutions, as well as generic extensions to existing passive SDN counter collection approaches.

#### 4.3.1.1 Distributed monitoring framework for SDN

Current SDN-frameworks are from a fault and performance management and monitoring perspective in general very limited, offering at most the possibility to measure flow statistics based on simple counters in the switches ( [42], [46], [43], [41]). To ensure telecom operator-scale for deployed UNIFY service platforms, it is crucial for the observability capabilities to be scalable and resource-efficient, since observability points can and will be instantiated in large numbers at various network locations, instructed to collect diverse measurements at high frequencies in order to meet carrier requirements such as failure-resilience, service-deployment flexibility, and dynamic scaling. We plan to advance state of the art of SDN fault and performance management tools in the following aspect:

- <u>RC1: Probabilistic in-network monitoring methods</u>: Efficient monitoring functions are crucial for enabling the UNIFY vision of dynamic and flexible service-chaining. This requires scalable and resource-efficient modelling of various performance indicators, as well as fast and flexible detection of the source of performance degradations with low level and high precision monitoring data, while minimizing control/data plane overhead. In UNIFY, we will develop scalable and adaptive approaches for performance monitoring, based on probabilistic methods for efficient calculation of statistical estimates of performance measurements for modelling of flow and link metrics, using distributed methods when applicable. To achieve scalability we aim to calculate the estimates as locally as possible, and provide the results as compact and expressive parametric distributions. The monitoring functions will be adaptive and controllable through high-level performance objectives, which will contribute to operational resource-efficiency and simplified configuration. In order to pursue these design goals, certain requirements on node resources need to be fulfilled for modelling of flow and link metrics (as listed in Section 5.1.1). These requirements include e.g. basic computational and storage facilities at the nodes for local estimation and aggregation, and additional counters in the data plane for obtaining richer statistics of the observed network behaviour at varying time scales. We believe that a distributed, probabilistic approach is absolutely necessary in order to deliver sufficiently precise data and predictions on monitored performance indicators with modest demands on infrastructure resources.

- <u>RC2: Scalable observability data transport and processing</u> : The data collected by observability points needs to be propagated to the control plane component of the virtual monitoring function, centrally located at the Controller Layer (controller). Typically, the data is aggregated to compute the final metrics required by the monitoring objective on this central location. However, this puts pressure on both the underlying network capacity, and also on the processing capabilities of the central aggregator. We will propose a method to optimize the processing and transport of observability data through the use of in-network aggregation. We will introduce in-network aggregation points (AP) to effectively reduce the load on both the network and the control plane component on the Controller Layer. APs will be deployed deeply into the network (e.g. as local CP components) in order to attract data from observability points in their proximity, aggregate and pre-process the data by a function specified by the monitoring objective, and send the data (with greatly reduced overhead) up to the control plane component. Here, one key problem will be finding the optimal placements of APs while trading-off between traffic load benefits and costs associated with instantiating APs. Existing work in this area, especially the VirtuCast algorithm [82], only consider commutative and associative aggregation or filtering functions. However, as monitoring functions are in general not commutative, the main research challenge is to extend the existing algorithmic framework of [82] to incorporate arbitrary monitoring functions, while still optimizing both AP placements and traffic usage. In summary, we plan to develop a scalable observability data transport and processing system that builds on the general ideas of VirtuCast and is adapted to the UNIFY architecture and Use-Cases.

### 4.3.1.2 Controller based performance monitoring for SDN

As noted in section 2.3.1.2 most of the performance monitoring solutions for Openflow networks operate under assumptions that are usually not met in a service provider scenario as envisioned in UNIFY: they assume re-active flow instantiation, triggered by the arrival of an unknown flow at a switch; and/or they assume fine-grained flow definitions with fully specified matching structures. Both of these assumptions are not met in most service provider networks, where operators typically will proactively populate flow tables during provisioning time (e.g. for routing purposes), and take advantage of the possibilities to aggregate traffic by coarse-grained flow definitions (i.e. by applying wildcards in certain fields of the flow definition table). Additionally, due to the inherently centralized nature of existing SDN control planes, controller based methods often suffer from the resulting scalability challenges with respect to signalling and notification load.

- <u>RC3: Low-overhead performance monitoring for SDN</u>: Motivated by the shortcomings of current solutions, it is our goal to design a scalable and programmable method that provides generic SDN performance data in a service provider scenario as depicted in UNIFY (e.g. use case in Section 3.1). In summary, the design goals are the following:

    • Accurate determination of link and flow performance (i.e. loss, delay, throughput)

    • Generic, technology independent solution based on the existing SDN components

    • Programmable interfaces to facilitate automation (e.g. for troubleshooting purposes)

    • Effectiveness against granularity of flow definitions and flow-deployment mode (reactive vs proactive)

    • Scalability with respect to signalling and data-plane overhead

We will investigate a method for estimating network performance metrics (at first focusing on packet loss and delay) which contributes to the observability capabilities of the network infrastructure in UNIFY [83]. The method will take advantage of user traffic transported using pre-deployed aggregated Openflow flow descriptors. Employing user traffic will reduce network overhead compared to most OAM tools that rely on active measurement methods, i.e. injection of packets or packet trains into the network. We will explore how to adapt existing ideas on low-cost monitoring (FlowSense [42]) to pre-provisioned and aggregated flow definitions in order to keep signalling overhead between control and data planes minimal. We will also investigate a programmable devolving mechanism of aggregated flow-definitions (defined with wildcards) into isolated microflows (with fully defined flow definitions) which can be used as measurement samples. This idea is inspired by DevoFlow [84], but requires extensions to support various performance metrics while being steered by policies on control and orchestration layers. Further ideas for potential extensions of the method include extensions for additional network performance metrics (such as throughput, packet reordering, etc.) and expandability toward non-network resources (i.e. integration with compute resources e.g. for a combined delay metric). With respect to deployment of the observability method, conditional activation of the capability will be considered, as well as consistent configuration across the network path in question.

#### 4.3.1.3 Passive measurement extensions to SDN

OpenFlow-enabled switches passively gather statistics in several places of the OpenFlow switch model, based on counting the packets and bytes traversing a particular part of a switch configuration. In the latest OpenFlow version (1.4.0 at the time of writing) these statistics are available for individual flow table entries, aggregate flow table entries, whole flow tables, group entries, meters, queues and ports. While these counter-based statistics provide a wealth of information about the system, the "rawness" of the measurements and the protocol used to access them makes using the data efficiently to calculate network metrics a difficult process with a large overhead. We are planning to extend various parts of the statistics system with capabilities on the switch to reduce overhead on both the network and the central controller.

The extensions we will investigate involve the inclusion of new metrics into existing structures, for example adding latency statistics to queuing and flow table statistics. Such metrics can easily be calculated locally on the switch, but require heavy overhead to calculate it at the centralized controller. Other extensions we will investigate are mechanisms for retrieval and use of existing statistics in a different manner. Such type of protocol messaging extensions can involve the inclusion of thresholds and associated alarms where today frequent polling would be necessary, as well as methods for obtaining for example the top 100 flow entries based on e.g. throughput rather than having to poll each individual installed flow table entry.

This work will be integrated as much as possible with both the distributed and controller-based monitoring approaches described above. The primary goal of the extensions is the proactive gathering of various metrics on link, queues, and nodes, such as their utilization and current latency, in order to support the orchestration and resource placement and optimization algorithms developed in WP3. At the moment the focus is on two extensions described below (RC4 and RC5), but this will be reconsidered if WP3 work shows the need for further metrics.

- RC4: Novel metrics in counter structures: Passive latency estimates could be obtained via monitoring the status of buffers in the network. Network buffers exists as a way to temporarily store bursts of incoming packets before transmitting them on an outgoing link, instead of dropping the packets. They are intended to perform this job transiently, quickly emptying the buffer after it has been used to store a burst. However, in many cases the buffers get filled up and stay full (obtaining a "standing queue"), losing their function of absorbing temporary bursts since they cannot buffer any more packets. Instead they just act as a source of latency in the network and cause additional packet drops. Active queue management (AQM) techniques, such as Random Early Detection, is intended to help mitigate this problem, known as "bufferbloat", but are notoriously difficult to configure correctly. Newer AQM such as Codel [85] measure the local standing queue and use it as a way of indicating when packets should be dropped. We believe the same metric gathered from multiple nodes could be used to predict the current latency over a path (for WP3, assisting in resource allocation) and provide insight as to the cause of network problems (as a troubleshooting tool). This requires extensions of existing statistics structures in OpenFlow.

- RC5: Efficient counter retrieval: Flow entry utilization (or "flow popularity") can be measured using OpenFlow statistics and used to indicate how to dynamically allocate and route flow entries in the network, in order to reduce overall latency, congestion, and network resource usage, by prioritizing the allocation of the most popular entries (which often conform to a Zipf distribution [86]). However, existing methods require continuous polling of all installed flow entries in order to determine the relative popularity of each flow. This is one of the problems that could be solved by adding new mechanisms in OpenFlow for retrieving statistics, by distributing the work of e.g. sorting the local flow entries based on a metric and periodically updating a controller with a small subset of entries and their statistics.

### 4.3.2 Verification

Enabling ongoing verification of code is an important goal of continuous integration as part of the DevOps concept. Here we mean verification with respect to the service definitions and configurations initiated by the *Service Developer*. Automated verification functions on each layer of the architecture facilitate verification as part of each step in the deployment process, allowing identification of problems early in the service lifecycle.

- RC6: Deploy-time functional verification of dynamic Service Graphs: The role of verification is key in the SDN scenario. A completely programmable network cannot disregard procedures to check the correctness of a network configuration set before it is deployed in the real system, especially if this comes from a user/customer with relatively low expertise. With this in mind, several tools have been proposed with the aim of enabling a (formal) verification of specific SDN configurations (e.g. with respect to availability of a path to the destination, absence of routing loops, access control policies, or isolation between virtual networks). Among the others, NICE [53], VeriFlow [65], and NetPlumber [66] represent significant examples in this direction (see Sections 2.3.3 and 2.3.4 for more details).

    The majority of these tools operate on network configuration rules (commonly OpenFlow), and in any case all of them do not consider active network functions (i.e. VNFs or middle-boxes that dynamically change the forwarding path of a flow according to local algorithms, as e.g. the IDS in the UC described in Section 3.1). In other words, these tools operate on the (centralized) programmability of the control plane only. This might be a limitation if we consider a possible network deploying active network functions, i.e. an environment that also enables programmability of the data plane in a distributed fashion. With this respect, novel tools for SDN verification are required, which could extend existing ones toward a fully programmable environment, such as one that includes network function virtualization.

    This also is particularly relevant in the context of the UNIFY project, which considers a network architecture made of virtual network functions running on physical network nodes. Tools like NICE, Veriflow or NetPlumber should be extended or adapted, in order to be able to work in this context too. Moreover, in the UNIFY architecture, network functions can be more than passive packet processing elements, simply combined into chains by means of proper forwarding rules set by a control plane level. For example, active network functions such as load balancers, packet marking modules, and intrusion detection systems might modify packet forwarding paths at run-time, or even the incoming traffic itself. For this reason, the adaptation of the standard SDN verification tools such as NICE and VeriFlow to the UNIFY architecture should be done not only to check the correctness of chain creation rules, but a more in depth verification of the resulting traffic paths should be done when active network functions are deployed into chains.

We will develop tools supporting the verification process during deployment of a Service Graph. For each step of the deployment process through the UNIFY architecture (from Service- to Infrastructure layer) we will investigate appropriate verification functions. Starting from a high-level verification of the customer input (i.e. the Service Graph), the verification process will then go more in depth in the chain configuration procedure by operating on the Network Function Forwarding Graph (NF-FG), which maps the Service Graph to the available network functions. If verification at this layer gives a positive feedback too, the tool will finally consider the low-layer information represented by the actual set of forwarding rules and other chain configuration parameters. Verification at the higher layers will be operated on graph descriptions and hence will mainly focus on topological properties (e.g., absence of forwarding loops or deadlocks). As this process involves lower layers of the architecture, it will have the opportunity to access additional information. This enables the verification of more quantitative properties (e.g. compliance with resource availability), as well as a more detailed and precise verification of the abovementioned topological ones. At the Infrastructure Layer, the tool will also handle possible active network functions, if any is deployed in the chain. This last verification step represents the most relevant advance beyond the current state of the art, as already mentioned. Here, the challenges stand in extending the already known verification techniques that operate on the control plane (e.g. on Openflow rules) so that virtual and active network functions can also be considered. This has to be done without a relevant degradation of verification times, so as to still allow fast SP-DevOps cycles. After the study of the already existing tools, our work will proceed with the design of adapted tools to the new context.

- <u>RC7: Run-time verification of forwarding configurations by enhanced ATPG</u>: The existing Automatic Test Packet Generation (ATPG) [51] tool for verification of deployed flow rules is very specific to one type of network and has many shortcomings and overheads. One of the overheads of ATPG is polling the network periodically for the forwarding state and performing all-pairs reachability. If this polling interval is significantly large, there are more chances that ATPG will have incorrect information of the network state as the state may change over time. If the polling interval is small, it brings additional load in the network. In addition, the current implementation of ATPG is only able to verify and test the action part of the forwarding rule. Therefore, the matching part of the forwarding rule remains untested. Furthermore, with the current ATPG implementation, only active rules can be tested: for example, an error in a backup rule cannot be detected in a normal operation or an error in the working path cannot be detected in a failure operation. Moreover, ATPG modifies a packet to add a history field: an ordered list of rules the packet matched so far in the network. ATPG uses this history field to localize the root cause of the issues.

The main goal of our enhanced test packet generation tool is to solve all the aforementioned limitations of ATPG by testing both matching part and action part and also test the inactive rules for protection paths. For this purpose, our tool will periodically send test packets in the network to verify its operational status and therefore, these errors can be reported automatically to the network engineers. For verification of data plane connectivity, it can verify the network state periodically and can report if errors are present.

### 4.3.3 Troubleshooting

Implementing the UNIFY architecture or a similar SDN architecture supporting traffic chaining necessarily leads to complex systems consisting of a large number of components. These components are located at several abstraction layers of the architectures. In order to identify the source of problems in such a complex and multi-layered environment, both *Service developers* and *VNF Developers* will require support of advanced and troubleshooting mechanisms which need to be available during development, deployment and operation phases of the SP-DevOps cycle. Troubleshooting tools will need to operate on several levels of the architecture and will largely take advantage of verification and observability tools introduced above (Sections 4.3.1 and 4.3.2), as well as complementing existing tools (Section 2.3). These tools will be used collectively for localizing the cause of a problem such as fault or performance degradation. By ruling out sources of the problem the actual root cause can be narrowed down.

- RC8: Automated troubleshooting workflows: The problem of integrated network troubleshooting was first described profoundly in [52]. In UNIFY, we will follow a similar approach that automatically addresses the whole architecture to debug Service Graph deployment, configuration and troubleshoot operational problems. For this, we will go beyond the state of the art in SDN debugging / troubleshooting at more than one area. First, we intend to design a debugging system for the multi-layered UNIFY architecture. Components of the multi-layer architecture will be able to define monitoring and debugging information accessible through a common interface. This interface will employ layer hopping logic for allowing the debugger to follow execution across layers and components. To enable such an automatic troubleshooting logic, we will consider interfaces towards the observability and verification capabilities developed in UNIFY, allowing, fine-grained and automatic on-demand control for integration with other fault and performance measurement mechanisms and tools.

- RC9: In-network troubleshooting: Empirical parameters for metric models, obtained via probabilistic in-network monitoring as described in Section 4.3.1.1, will be used for fault management and troubleshooting purposes. Working towards the objectives of increased scalability, automation and adaptability in a timely manner, we will investigate how typically centralized SDN monitoring could be augmented with distributed and autonomous approaches for in-network detection and localization of performance degradations to physical and logical network devices. For these purposes, we will consider change detection and event correlation methods, which will require e.g. resource-efficient information exchange between nodes and access to local log data stored in the nodes. The overall approach of detecting changes in local estimators enables early warning of potential degradations, which is an important aspect to address towards dynamic and flexible deployment and operation of service-chains.

- RC10: Troubleshooting with active measurement methods: Besides run-time verification of forwarding configurations, the automated test packet generation (ATPG) method can also be used to debug many errors dealing with forwarding rules, actions, links, bandwidth usage, latency, and network elements. While the core algorithm of ATPG is targeting the minimization of test-packet sets for maximal forwarding rule verification, the tool can also be used as active measurement method for observability metrics such as delay, bandwidth usage, loss etc. For troubleshooting, it can thus be integrated into an automated workflow to perform verification and active performance measurements.

### 4.3.4 VNF development support

The VNF Development process supports any *VNF developer* to conform to the DevOps principle "Develop and test against production-like systems". Developing, debugging, and deploying a VNF in a live system raises many architectural, performance, and security issues we plan to address. For example, in a multi-layer architecture like UNIFY, it is not trivial to identify lower-layer instances of a high-level service description, e.g., identify a running VNF process on a specific machine inferred from high-level service graphs. From the perspective of performance and security, the appropriate isolation is the most important issue to address in order to protect already configured and running service graphs. Most of these issues are highly dependent on parts of the UNIFY architecture developed in other WPs and will need to be studied in close cooperation, such as functionality related to deployment of VNFs within the orchestration layer (WP3) or actual instantiation on the infrastructure layer (WP5). For example, well-defined troubleshooting interfaces are required for software components implementing the functionalities of all UNIFY layers.

- <u>RC11: VNF development support</u>: We investigate in details and elaborate on several functionalities required by VNF developer that shed light on possible solutions. These include the following:

    o adding a new version of a VNF to the service catalogue (optionally with a blank resource model).

    o VNF-developer initiated queries for access points of instantiated network functions. These access points then can be used, for instance, to attach a debugger to the VNF under development.

    o automatically determining VNF resource model by creating a special service chain. This service chain consists of the VNF in question, necessary Observation Points, and possibly traffic sources.

    o upgrading an instantiated service graph on the fly with a new version of a VNF. This probably should result in gradually driving traffic to the new VNF as old flows terminate.

    o accessing a network debugger like break pointing tool from multilayer debugger. This multi-layer debugger concept enables the support/definition of novel troubleshooting mechanisms between arbitrary layers of the architecture.

# 5 Requirements for realizing SP-DevOps in UNIFY

This section presents initial requirements defined from the perspective of the SP-DevOps concept (detailed in Section 4), and is a revisited and extended version of the high level requirements towards the UNIFY architecture documented in Section 4.3 of D2.1 [3].

Most of the requirements and processes in the scope of UNIFY SP-DevOps will be centred on observability and troubleshooting of service chains and its building blocks. As we noted in Section 4.2.1, a monitoring function consist of one or several observability points (OP) instantiated on one or several nodes together with a control plane component for analysis and control of lower-level monitoring operations towards the observability points. In turn, an observability point operates in terms of a node-local control and data plane components for local analytics and measurement purposes.

The requirements fall into two main categories: The first (and larger part) of the requirements express technical demands on the UNIFY architecture, i.e., what the architecture should provide in order to meet the needs of the SP-DevOps concept (technical requirements, in short). Fulfilling the technical requirements is expected to be done in collaboration between WP4 and WP3 and WP5. The second part of the requirements formulates further guiding principles in terms of usefulness criteria from the users' point of view, which we call operational requirements. In Annex 2 we present a mapping of both technical and operational requirements onto the objectives of the Work Package as specified in the Description of Work.

It is important to note that the requirements documented below will be revised with technical concepts evolving in other technical work-packages and updates will be documented in forthcoming WP4 milestone reports and deliverables.

In the following, the key requirements are written in **bold**. Some of these requirements are of more general aspect and broken down to more detailed requirements.

## 5.1 Technical Requirements

Technical requirements are expressed towards both the infrastructure (node-level, NL) and towards higher layers of the UNIFY architecture, primarily the orchestration layer (OL). The requirement of the SP-DevOps concept toward the infrastructure level will explicitly target the Universal Node given that it is under development within the project and would thus be a better candidate for adding advanced programmable functionality, compared to other UNIFY infrastructure targets. The requirements towards the higher layers will put demands mainly on the orchestration layer and target all of its sub-layers defined in the overarching architecture, but most commonly the virtualized infrastructure manager (i.e. SDN controller) sub-layer.

### 5.1.1 Node-level (infrastructure) requirements

One of the key questions in terms of observability is how the UN will support observability components. This subsection collects the identified requirements on the Universal Node, which are a refinement of the initial monitoring requirements documented Section 4.5 of D5.1 [87], as well as Section 4.3 of D2.1 [3]

- **NL1: The UN must support advanced monitoring capabilities**

This high-level requirement corresponds to SP-DevOps Req. 3-2 in D2.1. With advanced monitoring capabilities we mean that observability points (OP) will go beyond collecting simple statistics and counters from physical and virtual infrastructure resources. As an example, it is envisioned that observability points provide scalable and detailed performance metrics. These components can in turn require support from the UNIFY architecture and infrastructure, which must provide means for the OPs to, for example, perform packet manipulation (e.g., adding timestamps, marking certain types of packets for monitoring, etc. as further detailed in NL4) and to implement simple decision functions on virtual infrastructure resources dedicated to the specific OP (thresholds for notifications, aggregation of events, etc.).

- NL1.1: Observability components must have access to read the resource state on physical and virtual levels and access to timestamped log data stored in infrastructure resources.

- NL1.2: Universal Nodes should provide required resources (e.g. memory, CPU, storage) resources for node-local analytics.

- NL1.3: Observability components must be able to access data required to derive basic network metrics (throughput, loss, delay, jitter) on different levels of granularity (packet, flow, links, etc.) on physical and virtual infrastructure resources (corresponds to D5.1,Monitoring Req. 1)

- NL1.4: Observability components must be allowed to perform node-local analytics in the UN, based on counters capable of performing arithmetic operations directly in the data plane (e.g. for the purpose of aggregation, filtering, etc.) (corresponds to D5.1,Monitoring Req. 9).

- NL1.5: Observability components must be able to apply sampling strategies on measurement data at different levels of granularity (packet, flow, etc.) (corresponds to D5.1,Monitoring Req. 4)

- **NL2: The UN must provide interfaces to access observability metrics of services and their associated components in a desired timely manner.**

This high-level requirement corresponds to SP-DevOps Req. 3-5 in D2.1. It should support dynamic service optimization by mandating interfaces that propagate monitoring information in various time-scales, including close to real-time. This means, for example, that the monitoring capabilities of the OpenFlow protocol are not sufficient:

counters in the data plane need to perform arithmetic operations allowing for modelling of flow counters at high sampling rates (i.e., less than 100 ms).

- NL2.1: Observability components must be able to regularly report observability data. Reporting should be done in user specified time intervals, or alternatively the interval can be controlled through deterministic/probabilistic limits or other conditions.

- **NL 3: Observability Points must allow for dynamic installation, activation and deactivation on request for service provisioning and operational aspects.**

The high level requirement NL3 corresponds to SP-DevOps Req. 3-6 in D2.1. It specifies that the UN needs to support dynamic activation and deactivation of observability components in order to realize conditional observability points.

- **NL4: The UN should provide monitoring information to higher layers in suitable level of detail and granularity through Observability Points.**

The high-level requirement NL4 corresponds to node-level aspect of SP-DevOps Req. 3-7 in D2.1. It implies that in order to provide information through OPs to higher-layers, the UN also needs to implement capabilities that can provide OPs with necessary low-level information in line with the previously listed requirements.

Both of the above two requirements (NL3 and NL4) demand interfaces to the Universal Node for interaction with its advanced capabilities. On the other hand, they necessitate providing advanced capabilities for the monitoring and troubleshooting tasks. The following list details these two aspects.

- NL4.1: Observability components must be able to perform packet manipulation (e.g., adding timestamps, mark a certain packet for monitoring, etc.) (corresponds to D5.1, Monitoring Req.3)
- NL4.2: Observability components should be able to dynamically instantiate software-defined counters executed directly in the data plane (corresponds to D5.1, Monitoring Req.9).
- NL4.3: Universal Nodes should support instantiation of additional counters in the data plane as it enables modelling of counter values observed with fine granularity.
- NL4.4: Universal Nodes should allow for implementation of simple decision functions on virtual switches (thresholds for notifications, aggregation of events, etc.)
- NL4.5: Observability components must be able to perform active and passive measurements to observe and model different aspects of the network behaviour for monitoring and troubleshooting purposes. The use of active measurements means that observability-defined packets would need to be created within a Universal Node and inserted in the datapath such that they share the faith of the flows under test.
- NL4.6: Observability components and/or the infrastructure management layer must be able to instantiate, maintain and update sets of counters on virtual infrastructure resource.

- NL4.7: Observability components should also be able to instantiate, maintain and update more complex structures (e.g., Arrays) instead of single values for counters (corresponds to D5.1,Monitoring Req.2)

- NL4.8: Observability components need to exchange messages directly between physical and virtual infrastructure resources for efficient use and re-use of monitoring data as well as for troubleshooting purposes. However, observability components should not exchange information directly when they are situated in different administrative domains, for example customer VNFs "under observation" shall not receive information directly from observability components belonging to the infrastructure provider.

### 5.1.2 Orchestration level requirements

This subsection collects the identified requirements on the higher layers of the UNIFY architecture, primarily the orchestration layer, but also aspects of the service layer. The aim of these requirements is to outline the consequences and associated SP-DevOps requirements that arise given the overarching key requirements from D2.1.

- **OL1: The UNIFY architecture must support capabilities to develop and test components.**

This high-level requirement corresponds to SP-DevOps Req. 3-1 in D2.1. It aims to facilitate increased service velocity towards customers by continuous deployment and integration practices. This reflects DevOps principles by making it possible to develop and test VNFs and NF-FGs in production-like systems. These capabilities can for example include creation and isolation of resources slices (computer, network, storage) and isolation or special treatment of downstream traffic from development virtual network functions (VNF). The architecture must also be capable of running different versions of the same VNF at the same time.

- **OL2: The UNIFY architecture must support automated integration of monitoring, troubleshooting and verification capabilities.**

This high-level requirement corresponds to SP-DevOps Req. 3-3 in D2.1. It mandates programmable interfaces towards monitoring and verification capabilities on all architecture layers to support automation of operational processes. Automation of operational processes will allow adaptation and coordination of workflows for service design, development and operations teams.

- **OL3: The UNIFY architecture should be able to react accordingly to reports and notifications generated by observability and verification components.**

This high-level requirement corresponds to SP-DevOps Req. 3-4 in D2.1. Since observability and verification components will asynchronously generate reports and notifications (e.g., exceptions with respect to pre-configured or adaptively set thresholds and limits, such as changes, performance degradations, SLO breaches, etc), the above requirement specifies that the UNIFY orchestration and service layers will need to provide means to react in a way to mitigate the reported problem. This could be, e.g., to forward the notification through a service layer interface to the application logic, or to trigger an appropriate function within the orchestration or service layer (e.g., a scaling function, or a function providing some sort of re-optimization through changes in VNF placement or rerouting).

- OL3.1: The OL must provide parameter updates and other information relevant for (re-) configuration of monitoring functions upon operational changes in the network and service-chains, in order to obtain accurate monitoring information and maintain the capability of making near-optimal management decisions in a timely manner.

- OL3.2: The OL should include deployment of monitoring functions within the deployment of a service chain. This includes deployment of virtual monitoring functions (including all necessary OPs) as well as configuration input on the appropriate level of detail for the specific network functions (e.g. KQIs translated to KPIs, etc.)

- OL 3.3: Management-related functional components in the architecture should provide an observability components operating in the physical/virtual devices with relevant timestamped historical data, logs, events and other data (such as topology) upon request or as part of instantiating an observability component (e.g. topology).

- OL3.4: All components within UNIFY architecture should report all errors in each round of tests

- OL3.5: The OL must provide observability components with specified conditions for dynamic activation and deactivation of OP operations (allowing for conditional OPs).

- **OL4: The UNIFY architecture and its components should provide monitoring information in suitable level of detail and granularity according to the needs of applications or functional blocks within the architecture.**

This high-level requirement corresponds to SP-DevOps Req. 3-7 in D2.1. Here, the demand for appropriate information for different abstraction and virtualisation needs is covered. For example, information might be needed per application, user or even fine-grained element in the architecture.

- OL4.1: Monitoring functions and observability components must be able to send reports and asynchronous notifications to the Virtualized Infrastructure Management Layer (e.g., exceptions with respect to pre-configured or adaptively set thresholds and limits, such as changes, performance degradations, SLO breaches, etc)

- OL4.2: The UNIFY architecture should contain analytics functions on all architectural layers capable of aggregating different types of metrics collected and pre-aggregated by multiple components into domain or service chain wide KPIs

- OL4.3: Management functions and observability components must be allowed to perform in-network aggregation and processing in the infrastructure layer, in order to produce monitoring information at the specified detail and granularity in a resource efficient and timely manner.

- OL5: The UNIFY architecture must support automated verification of services and their formal representations.

This high-level requirement corresponds to SP-DevOps Req. 3-8 in D2.1. Verification tools offer possibilities to debug service components during design time to ensure intended functionality, as well as ensuring resource availability and verification of parameters settings during deployment time based on the specifications of service graphs and components. During deployment time, the validation and verification tools may need to operate within specified bounds (e.g., time limits) to ensure rapid service-chain deployment or re-deployment.

The interfaces supporting automated verification of services and forwarding graphs of the above requirement must work on all architectural layers (i.e., levels of abstraction), e.g., formal verification of definitions and configurations with respect to validity and inconsistency. To support verification functions, relevant pieces of information must be provided by the UNIFY architecture.

- OL5.1: The network operator must be able to verify the successful deployment and operation of virtual functions and service graphs consisting thereof.

- OL5.2: Verification of the operation of different paths in the service graph must be supported dynamically and on-the-fly.

- OL5.3: Verification module at Service Layer must receive (or be able to access) information about the service graph representing the service graphs to be verified and other related parameters, e.g., the ordered list of VNFs in a service graph.

- OL5.4: Verification module at Orchestration Layer must receive (or be able to access) information about the NF-FG representing the service graphs to be verified and a description of the underlying resources and topology.

- OL5.5: Verification module at controller layer must receive (or be able to access) information about the overlay topology of the service graphs to be verified and a description of the available computing and storage resources and of the VNF chain configuration rules.

- OL5.6: Verification modules must receive (or be able to access) possible verification policies to verify for a specific service graph (e.g., "all packets must pass through a given firewall").

- OL6: The verification functions should operate at design and (re-)deployment time.

This high-level requirement corresponds to SP-DevOps Req. 3-9 in D2.1. It is in line with the verification module described previously, but it additionally stresses the importance its place within the different workflows the users interacts with the UNIFY architecture.

## 5.2 Operational Requirements

This last part of the requirements formulates further guiding principles from the operators point of view, which we refer to as operational requirements. In part, these requirements reflect back on the SP-DevOps concept itself and will be taken into account during method development and interface specification. They are relevant in particular for the interaction with WP3 and our contributions to the integrated prototype.

- OR1: The impact of monitoring, verification and troubleshooting on the performance of the UNIFY architecture should be as low as possible.

The scalability requirements can be broken down at many levels. For example, one aspect is the efficiency of the measurement data transport. For the purpose of resource-efficiency, scalability and timeliness, monitoring functions based on in-network aggregation and processing in certain types of observability components are needed. For measurement intensive monitoring applications, the transport of measurement data collected by observation components needs to be efficient in the use of bandwidth. In-network processing also provides increased scalability in the processing of large amounts of fine-grained measurement data as the computational load is distributed. The degree to which in-network monitoring function can be instantiated during deployment and operation in the UNIFY framework is a trade-off between the type of observability needed, level of detail desired and available resources.

- OR2: The proposed SP-DevOps methods and tools must be in harmony with existing operational processes.

For authentication, authorization and accounting purposes, the operator must be capable to validate the type/class/quality of service that a customer receives. Accountability is also important for the sake of SLAs monitoring and billing processes, but also to avoid customers' misuse or malicious behaviour. Additionally, the network operator must be able to verify the successful development and operation of virtual functions and service chains consisting thereof. Moreover, the verification of the operations of different paths in the service graph must be supported dynamically and on-the-fly. The network operator should be able to dynamically resolve/troubleshoot issues with VNFs or service graphs and in a flexible manner, which also includes the possibility to temporarily start, stop, configure and deploy observability points as necessary. Automatic monitoring and troubleshooting processes could be employed (even periodically), instead of manual or ad-hoc procedures that involves human intervention.

# 6 Conclusions

In this deliverable, we present a sketch of the SP-DevOps concept as a proposal that would enable telecom providers to support faster introduction of new services. We identified characteristics that differentiate DevOps in a telecom provider environment compared to the datacentre, where the concept originates. We identified two developer roles – the Service Developer defines a service graph associated to a new service, and the VNF developer implements the software associated with a new virtual network function and maintains existing ones. We also identified the need for an Operator role that, in addition to the regular daily operation duties from the eTOM and ITIL frameworks, empowers the developers to troubleshoot problems with their code in conditions close to a real production environment. In our definition, SP-DevOps reposes on three pillars aligned towards the Work Package objectives specified in the Description of Work: observability, troubleshooting and verification. Processes associated to each one of these areas were described, pinpointing components of the functional architecture that would be involved in the implementation.

We outlined the fact that the concept of SP-DevOps is wider than WP4 and requires cooperation with WP3 and WP5. In this respect, we formulated a set of technical requirements towards the programmability framework, the Universal Node as well as towards the overall architecture developed in WP2. WP4 will also have a role in fulfilling parts of the technical requirements. These requirements are complemented by two generic operational requirements aimed at the observability, verification and troubleshooting areas to be approached in this Work Package.

We surveyed the state of the art for the areas of software-defined infrastructure monitoring, verification and troubleshooting, with a focus on software-defined networks. Several key issues were identified and preliminary ways of addressing them in the duration of the project were outlined. Monitoring and troubleshooting approaches were found to be highly limited in terms of the level observability that can be efficiently provided in a UNIFY context. We will address this by designing resource-efficient, scalable and controllable methods capable of operating dynamically relative to specified KPIs and adapt to performance changes in service graphs. This will include development of more advanced methods based on capabilities of the OpenFlow standard as well as extended OAM mechanisms for obtaining richer information about the virtualized infrastructure behaviour while lowering the operational overhead. We discussed the need for and presented a definition for Observability points, which are virtual network functions that implement advanced monitoring functionality. Verification approaches assume a centralized and statically-controlled network configuration, which is different from the dynamic UNIFY production environment. We will design verification mechanisms that exploits the programmability of the UNIFY environment to provide dynamic adaptation to service graphs deployments and runtime evolution. The VNF Development framework provides the service developer the necessary means for testing, debugging and deploying services, and relies partially on the monitoring, troubleshooting, and verification processes. Main research challenges to be addressed include efficient tracking and identification of different VNF processes common to a service graph running in the UNIFY multi-layer architecture, as well as maintaining the specified performance of service-chains. The research challenges we outlined cover all the objectives proposed in the Work Package description document.

The following steps will be taken to advance the SP-DevOps sketch towards an initial concept that will be detailed in MS4.1. We will identify and specify interfaces associated to passing SP-DevOps-relevant information between components of the functional architecture. In cooperation with the service instantiation and deployment framework developed in WP3, we will work on specifying how to describe monitoring and verification capabilities such that they could be integrated in the UNIFY production environment. Together with the work on the infrastructure and hardware aspects in WP5, we will work on further understanding how the Universal Node can support our requirements for programmable monitoring capabilities. The progress of these discussions, along with the progress made by partners in designing monitoring, verification and troubleshooting capabilities will be reported in MS4.1. As the map of capabilities and interfaces becomes more complete, we will start documenting how an integrated WP4 prototype could be built for demonstrating the advantages of our combined approaches. The progress in this direction will too be documented in MS4.1.

# Annex 1 : Detailed service configuration steps

**Example #1: VPN service**

In Figure 24, an MPLS VPN network is depicted.

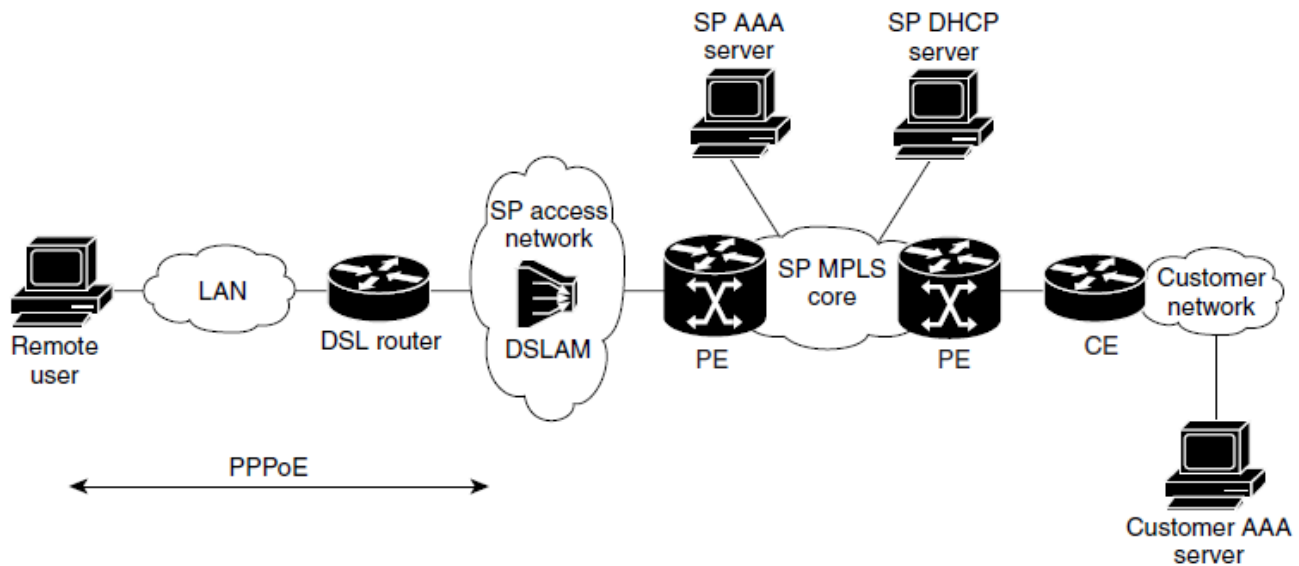

*Figure 24: MPLS VPN Network [20]*

Below, we present the tasks that need to be completed by a network operator to configure, monitor and maintain, and trouble-shoot such an MPLS VPN network.

### A. Configuration

1. Configuring the MPLS Core Network
    a. Enabling Label Switching of IP Packets on Interfaces
    b. Configuring Virtual Routing and Forwarding Instances
    c. Associating VRFs
    d. Configuring Multiprotocol BGP PE to PE Routing Sessions
2. Configuring Access Protocols and Connections
    e. Configuring a Virtual Template Interface
    f. Configuring PPP (or PPPoE) over ATM Virtual Connections and Applying Virtual Templates
3. Configuring and Associating Virtual Private Networks
    g. Creating a VRF Configuration for a VPN
    h. Associating a VRF Configuration for a VPN with a Virtual Template Interface
4. Configuring RADIUS User Profiles for RADIUS-Based AAA

5. Verification

- Configuration of PPPoE to MPLS VPN
    - Drawbacks: manual configuration, complexity
    - Number of commands in CLI: 183
    - Probability of mistake: Very high

B. **Monitoring and maintenance**

1. Monitoring and maintenance of the MPLS Configuration
    a. Verification of Successful Running of the Routing Protocol
    b. Verification of MPLS
    c. Verification of Connections Between Neighbors
    d. Verification of Label Distribution
    e. Verification of Label Bindings
    f. Verification of Labels Are Set
2. Monitoring and maintenance of the MPLS VPN
    a. Verification of VRF Configurations
    b. Verification of the Routing Table
    c. Verification of PE to PE Routing Protocols
    d. Verification of PE to CE Routing Protocols
    e. Verification of the MPLS VPN Labels
    f. Testing the VRF
- Monitoring and maintenance of PPPoE to MPLS VPN
    - Drawbacks: manual, high frequency, prone to mistakes

**Example #2:IPTV service**

A. **IPTV network management drawbacks/challenges**

- Multi-vendor equipment such as head-end equipment, middleboxes, VoD servers, CAS/DRM equipment, etc.
- Careful configuration of multiple systems and network devices; necessary to avoid post-installation issues.
- Critical monitoring of QoS across the network, i.e. from head-end to the access network.
- Complex trouble-shooting and isolation of problems.
- Constant capacity monitoring and instant switch to an alternative (back-up) path to carry the video.

B. **Service provisioning – tasks to be fulfilled**

1. Service activation: CRM, DSLAM EMS's, subscriber management, CPE provisioning system, identity and access control (AAA servers), billing system for service enablement.
2. Configuration management.
3. Service assurance: execution of proactive and reactive maintenance to ensure that the IPTV service performs according to the QoS levels defined in SLAs.
4. Fault management.

C. **Network management and maintenance**

1. VoD monitoring
2. Video quality monitoring
3. Performance management
    - IPTV service KPIs: Packet loss, jitter, latency, channel change time
    - Device KPIs: CPU, memory, buffer utilization
    - Network KPIs: CIR utilization, queue drops, dropped frames
    - Methodology: probes, device instrumentation

# Annex 2 : Mapping onto WP4 Objectives

*Table 1: Research Challenges described in D4.1, mapped on the WP4 Objectives in the DoW*

| DoW WP4 Objective | Research Challenges |
|---|---|
| O4.1 Evaluate and demonstrate, in an agile manner, the SP-DevOps concept for selected scenarios, including the development of the Service Provider DevOps prototype (DevOpsPro) | N/A |
| O4.2 Define conditional observability points located on Universal Nodes and develop an automated approach for deploying them consistently | Partly addressed by RC1, RC2, RC3, RC8, RC9, RC10 |
| O4.3 Develop scalable service monitoring approaches, adapted to software-defined networks, that are efficient in reducing the number of manual diagnosing steps and amount of observation data transiting on the network | RC1: Probabilistic in-network monitoring methods<br><br>RC2: Scalable observability data transport and processing<br><br>RC3: Low-overhead performance monitoring for SDN<br><br>RC4: Novel metrics in counter structures<br><br>RC5: Efficient counter retrieval<br><br>RC9: In-network troubleshooting<br><br>RC10: Troubleshooting with active measurement methods |
| O4.4 Design methods for verifying service chain functionality at runtime and locating service chain faults | RC7: Run-time verification of forwarding configurations by enhanced ATPG |
| O4.5 Enable automatic definition of workflows for verification and activation tests for dynamic service chains | RC8: Automated troubleshooting workflows<br><br>RC11: VNF development support |
| O4.6 Enable the possibility to verify service chains within the limit of one development cycle | RC6: Deploy-time functional verification of dynamic Service Graphs |

80                                Deliverable D4.1                                10.02.2015

Table 2: High-level requirements described in D4.1, mapped on the WP4 Objectives in the DoW

| DoW WP4 Objective | High-level requirements |
|---|---|
| O4.1 Evaluate and demonstrate, in an agile manner, the SP-DevOps concept for selected scenarios, including the development of the Service Provider DevOps prototype (DevOpsPro) | N/A |
| O4.2 Define conditional observability points located on Universal Nodes and develop an automated approach for deploying them consistently | NL1: The UN must support advanced monitoring capabilities<br><br>NL 3: Observability points must allow for dynamic installation, activation and deactivation on request for service provisioning and operation aspects<br><br>NL4: The UN should provide monitoring information to higher layers in suitable level of detail and granularity through Observability Points |
| O4.3 Develop scalable service monitoring approaches, adapted to software-defined networks, that are efficient in reducing the number of manual diagnosing steps and amount of observation data transiting on the network | NL1: The UN must support advanced monitoring capabilities<br><br>NL2: The UN must provide interfaces to access observability metrics of services and their associated components in a desired timely manner<br><br>NL4: The UN should provide monitoring information to higher layers in suitable level of detail and granularity through Observability Points<br><br>OL4: The UNIFY architecture and its components should provide monitoring information in suitable level of detail and granularity according to the needs of applications or functional blocks within the architecture<br><br>OR1: The impact of monitoring, verification and troubleshooting on the performance of the UNIFY architecture should be as low as possible<br><br>OR2: The proposed SP-DevOps methods and tools must be in harmony with existing operational processes |
| O4.4 Design methods for verifying service chain functionality at runtime and locating service chain faults | OL1: The UNIFY architecture must support capabilities to develop and test components<br><br>OL5: The UNIFY architecture must support automated verification of services and representations<br><br>OR2: The proposed SP-DevOps methods and tools must be in harmony with existing operational processes |
| O4.5 Enable automatic definition of workflows for verification and activation tests for dynamic service chains | OL1: The UNIFY architecture must support capabilities to develop and test components<br><br>OL2: The UNIFY architecture must support automated integration of monitoring, trouble-shooting and verification capabilities |
| O4.6 Enable the possibility to verify service chains within the limit of one development cycle | OL5: The UNIFY architecture must support automated verification of services and representations<br><br>OL6: The verification functions should operate at design and (re-)deployment time.<br><br>OL3: The UNIFY architecture should be able to react accordingly to reports and notifications generated by observability and verification components |